\documentclass{article}
\usepackage[utf8]{inputenc}
\usepackage[margin=2cm]{geometry}

\usepackage{caption}
\usepackage{subcaption}

\usepackage[
    backend=biber,
    sorting=none
    ]{biblatex}
    
\usepackage{graphicx} 
\graphicspath{ {figures/} } 

\usepackage{comment}

\usepackage{amsmath} 
\usepackage{amssymb}

\usepackage[super]{nth}

\begin{document}

\title{A Conversational Brain-Artificial Intelligence Interface} \author{Anja Meunier$^{*,\dagger,1,2}$, Michal Robert Žák$^{\dagger, 1}$, Lucas Munz$^1$, \\  Sofiya Garkot$^1$, Manuel Eder$^1$, Jiachen Xu$^{1,2}$, and Moritz Grosse-Wentrup$^{1,3,4}$\\
\ \\
\small{$^1$Research Group Neuroinformatics, Faculty of Computer Science, University of Vienna, Vienna, Austria}\\
\small{$^2$Doctoral School Computer Science, Faculty of Computer Science, University of Vienna, Vienna, Austria}\\
\small{$^3$Vienna Cognitive Science Hub, University of Vienna, Vienna, Austria}\\
\small{$^4$Research Network Data Science, University of Vienna, Vienna, Austria}\\
\small{$^\dagger$These authors contributed equally to this work.}\\
\small{*anja.meunier@univie.ac.at}
}
\date{February 26, 2024}

\maketitle

\begin{abstract}
We introduce Brain-Artificial Intelligence Interfaces (BAIs) as a new class of Brain-Computer Interfaces (BCIs). Unlike conventional BCIs, which rely on intact cognitive capabilities, BAIs leverage the power of artificial intelligence to replace parts of the neuro-cognitive processing pipeline. BAIs allow users to accomplish complex tasks by providing high-level intentions, while a pre-trained AI agent determines low-level details. This approach enlarges the target audience of BCIs to individuals with cognitive impairments, a population often excluded from the benefits of conventional BCIs. We present the general concept of BAIs and illustrate the potential of this new approach with a Conversational BAI based on EEG. In particular, we show in an experiment with simulated phone conversations that the Conversational BAI enables complex communication without the need to generate language. Our work thus demonstrates, for the first time, the ability of a speech neuroprosthesis to enable fluent communication in realistic scenarios with non-invasive technologies.
\end{abstract}

\section{Brain-Artificial Intelligence Interfaces}
Brain-Computer Interfaces (BCIs) are systems that \textit{give their users communication and control channels that do not depend on the brain’s normal output channels of peripheral nerves and muscles} \autocite{wolpaw2000brain}. BCIs are used in a variety of contexts, e.g., to provide a communication channel to severely paralyzed patients \autocite{birbaumer2000thought}, to adapt human-computer interaction systems to their users' mental states \autocite{zander2011towards}, and to support post-stroke motor rehabilitation \autocite{grosse2011using, gomez2011closing}. When used as assistive communication tools, BCIs typically build on intact cortical processes to generate the command signals. In other words, they presume that the user is cognitively capable of executing the desired task, and the BCI merely bridges the severed connection between the brain and the outside world. However, more than $40\%$ of first-ever stroke patients, a group that arguably stands to benefit tremendously from assistive technologies, exhibit cognitive impairments \autocite{lawrence2001estimates} and about one-third of stroke patients show symptoms of aphasia, i.e., deficits in language expression and comprehension \autocite{dickey2010incidence}. These deficits often result from extensive cortical brain damage. Patients with extensive cortical damage are unlikely to benefit from BCI systems that attempt to read out information from brain areas that have been directly affected or even destroyed by the stroke. For example, a stroke patient with a middle cerebral artery (MCA) infarct is unlikely to benefit from a BCI that attempts to reconstruct speech by decoding movements of vocal tract muscles from the laryngeal motor cortex, a cortical area typically affected by such an infarct. To build assistive communication devices for patients with extensive cortical damage and cognitive deficits, we must consider the feasibility of using neurotechnology to \textit{replace} lost cognitive functions.

Due to tremendous advances in artificial intelligence (AI) in the past decade, AI systems trained on enormous amounts of data are now able to perform a broad range of cognitive tasks on a level similar to or even exceeding human performance, e.g., in image- \autocite{russakovsky2015imagenet} and speech recognition \autocite{deng2013new}, strategic planning \autocite{mnih2015human, meta2022human}, and language generation \autocite{gpt3}. By using existing BCI technologies to interface with such AI systems, they can be leveraged to perform cognitive tasks that the patient has lost. We term this new class of systems Brain-Artificial Intelligence Interfaces (BAIs) to distinguish them from classical BCIs that merely attempt to (re-)connect a fully functional mind with the outside world. 

By delegating some parts of the cortical processing pipeline to an AI agent, BAIs replicate the natural division of responsibilities when performing complex cognitive tasks. Humans typically control higher-level functions deliberately, while lower-level processes are automated. When speaking, for example, the exact choice of words is automated to a large extent in natural conversation, while the content of the produced speech underlies executive control \autocite{Levelt89Speaking, Mayer2006Automaticity}. Similarly, in a BAI, the user's intentions are fed into the AI system, thus steering it in a particular direction, while the details of the response are determined by the AI system's extensive experience in the given task. Thus, the BAI user controls the high-level intention, while a pre-trained AI agent automatically fills in the low-level details.

The idea of feeding a BCI user's high-level intentions to an automated system that determines lower-level details has already been successfully applied in the area of robotic arm control \cite{lampe2014, downey2016blending, muelling2017autonomy, XU2019121, cao2021brain}. As artificial intelligence capabilities continue to grow, we anticipate that BCIs that rely on sharing control between users and AI agents will become increasingly sophisticated and widespread throughout established and emerging areas of BCI usage. We, therefore, argue that an overarching and general characterization of this new class of BCI systems beyond motor neuroprostheses is in order. 

\begin{figure}[htb!]
    \centering
    \includegraphics[width=0.55\textwidth]{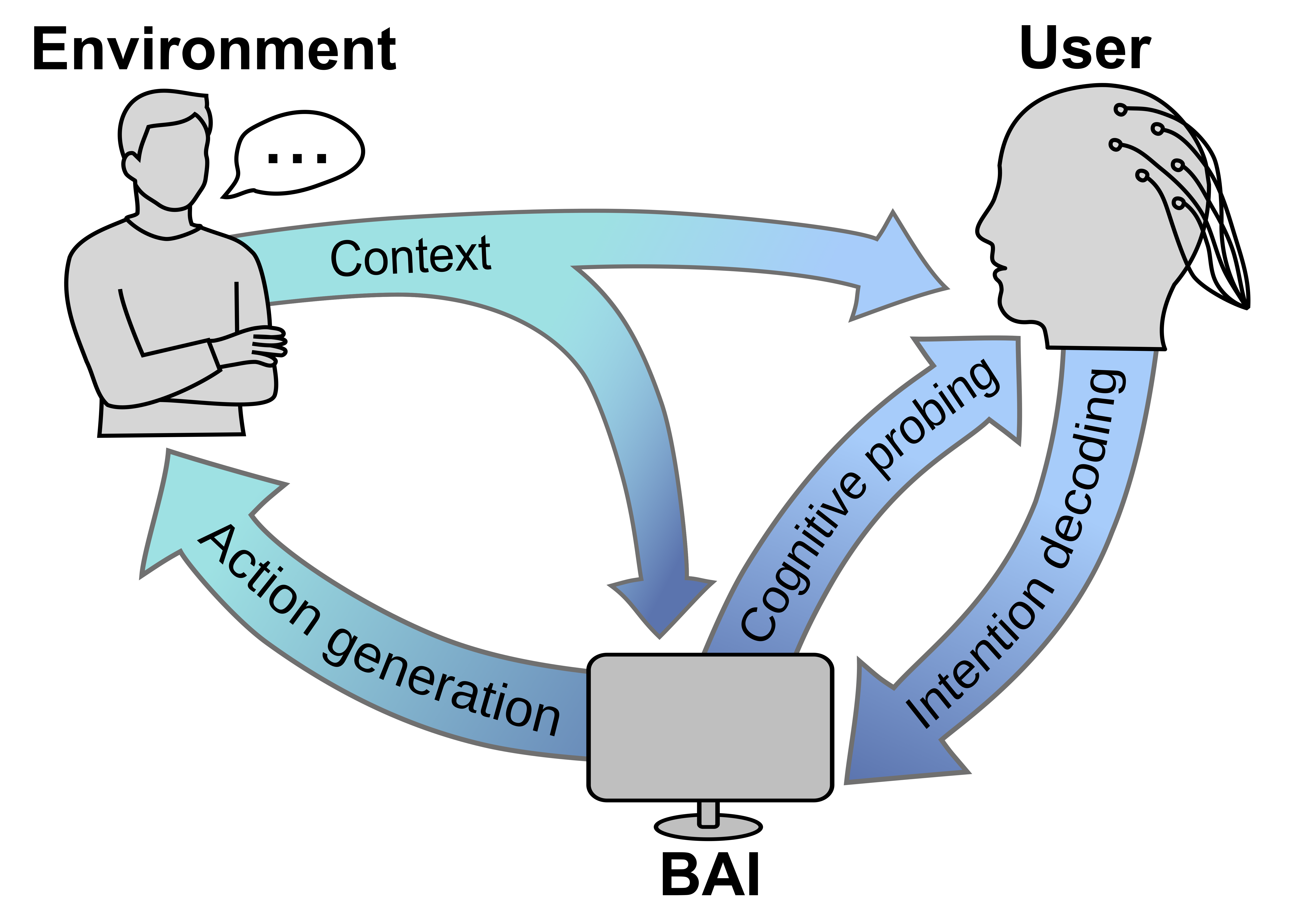}
    \caption{Structure of a Brain-AI Interface. The BAI first receives contextual information from the environment and then probes the user about their intentions. These are subsequently decoded from neurophysiological data and then translated into action by the BAI.}
    \label{fig:baistruct}
\end{figure}

The generic structure of and information flow in a BAI are shown in Figure \ref{fig:baistruct}. Its primary components are contextual input, cognitive probing \cite{krol2018towards, krol2020cognitive}, intention decoding, and action generation. The contextual input component represents how the BAI receives knowledge about the surroundings that the BAI user is experiencing and wants to react to. This is important because behavior always occurs within a contextual environment, and knowledge of this context provides information about appropriate reactions. For example, questions by a conversation partner represent relevant contextual information, which must be supplied to a Conversational BAI as input. After the BAI has gained information about the context the user wants to react to, it must gather information about the user's intentions. It probes the user for relevant information, and the intentions of the user are subsequently decoded from the user's neurophysiological data. The chosen cognitive probing approach and the selected method of intention decoding must be closely integrated. In the ideal, most intuitive case, whenever the BAI needs additional information to generate an action, it probes the user by directly decoding their intentions from the neurophysiological data. The challenges associated with this optimal approach, which stem from the current scarcity of suitably specialized AI agents and the complexity of decoding user intentions directly, are explored below with an emphasis on Conversational BAIs. The final step in closing the loop of the interaction between the outside world and the user is action generation. After identifying the user's intention, the BAI generates an action that best matches the user's desired action. As a result, the BAI user gains the capability to execute their intentions and influence their surroundings through the BAI, effectively exercising agency and closing the feedback loop with the environment.

In the following, we demonstrate the power of this novel class of systems in a Conversational BAI, a class of BAIs designed to aid users with severe language impairments. We evaluate our system in an experiment with simulated phone conversations and demonstrate the efficacy of Conversational BAIs for accomplishing everyday communication tasks. 

\section{Conversational BAIs}
The importance of language as a means of human communication cannot be overstated. Its loss can be a profoundly life-altering experience with significant implications for an individual's mental health and overall quality of life \cite{cruice2003finding}. Recent years have witnessed remarkable advancements in language-related artificial intelligence, primarily owing to large language models (LLMs) based on transformer architectures \cite{vaswani2017attention, gpt3, gpt4, touvron2023llama}. These models now stand at the forefront of AI development, capable of generating human-level language. Our goal is to leverage this ability for the development of intuitive neuroprostheses for patients with severe language impairments: Conversational BAIs.

Conversational BAIs empower users to communicate verbally by relying on a sophisticated conversational agent to handle the intricacies of language generation. When presented with a question as contextual input, the conversational agent probes the user for the general intent of their response and subsequently generates grammatically sound, contextually appropriate responses that align with the user's intention.

One of the central challenges in developing Conversational BAIs revolves around selecting and adapting the conversational agent, as BAIs rely heavily on the quality of the AI agent at their core. State-of-the-art LLMs are trained on extensive corpora of natural language data, and conversational agents are typically fine-tuned on conversation data \autocite{gpt3, dialogpt, madotto2020plug}. They generate responses by randomly selecting from a learned distribution of words or sub-words, conditioned on the input prompt and previously generated tokens. However, such systems are not constrained to produce a particular intention in their answers. For example, when prompted, "How is the weather today?" two plausible answers may be "It is very nice and sunny today." or "Today it is cold and cloudy." State-of-the-art conversational agents are trained to produce coherent and probable answers but are not specialized to receive information about and subsequently represent a particular viewpoint in a conversation. Ultimately, developing a new class of conversational agents optimized to speak \textit{for}, rather than \textit{with} someone, is crucial to reaching the full potential of Conversational BAIs.

Another critical aspect of Conversational BAI design involves determining the optimal point in the neuronal processing pipeline for decoding the user's intentions. An ideal decoding approach considers how much control over the details of the response the BAI user can still retain on their own.  This highly depends on which cortical processes remain intact. For instance, an individual with anarthria, the complete inability to perform speech movements, for example, due to a stroke or amyotrophic lateral sclerosis (ALS), is typically still able to generate imagined speech. In recent years, complex decoding methods such as sequence-to-sequence neural networks have led to remarkable results in speech decoding. Vocalized, imagined, or attempted speech can be synthesized from intracortical microelectrode array data \autocite{metzger2023high, willett2023high} or ECoG recordings \autocite{angrick2019, anumanchipalli2019, proix2022imagined}, enabling severely paralyzed patients to communicate. However, this approach fails for patients who struggle to come up with the words to express their mental concepts, in particular stroke patients with aphasia. 

Therefore, developing novel specialized decoding methods for this patient group is paramount. In an ideal scenario, this means directly decoding user intentions much earlier in the cortical processing pipeline. However, this requires sufficiently granular neurophysiological data and an in-depth understanding of the cortical language processing pipeline. How mental concepts are represented in the early stages of cortical language processing and whether aphasic patients' speech intentions can be decoded from such neuronal representations is yet unclear and remains an open area of research.

As an alternative non-invasive, less direct method for cognitive probing and decoding, we propose and test an approach where the conversational agent generates multiple suggestions in response to the contextual input, with the BAI user selecting among them using classic BCI paradigms, for example, code-modulated visually evoked potentials \cite{nagel2019plos}. Selecting from the suggested responses allows the user to express a wide range of concepts, even when more complex neurophysiological signals are unavailable. 

In the following, we introduce the first Conversational BAI based on EEG and present its evaluation in simulated phone conversations. We describe its components (Section \ref{sec:prototyp}), the experimental design (Section \ref{sec:expdesign}), and present the results of our analysis (Section~\ref{sec:analysis}).

\subsection{EEGChat: A non-invasive Conversational BAI}
\label{sec:prototyp}

EEGChat is a proof-of-concept Conversational BAI demonstrating the possibilities of this technology. Its versatility allows users to engage in conversations on any topic via an indirect cognitive probing approach based on non-invasive EEG recordings. A demonstration video showcasing it in action can be found in the supplementary materials. 

\begin{figure}[htb]
    \centering
    \includegraphics[width=\textwidth]{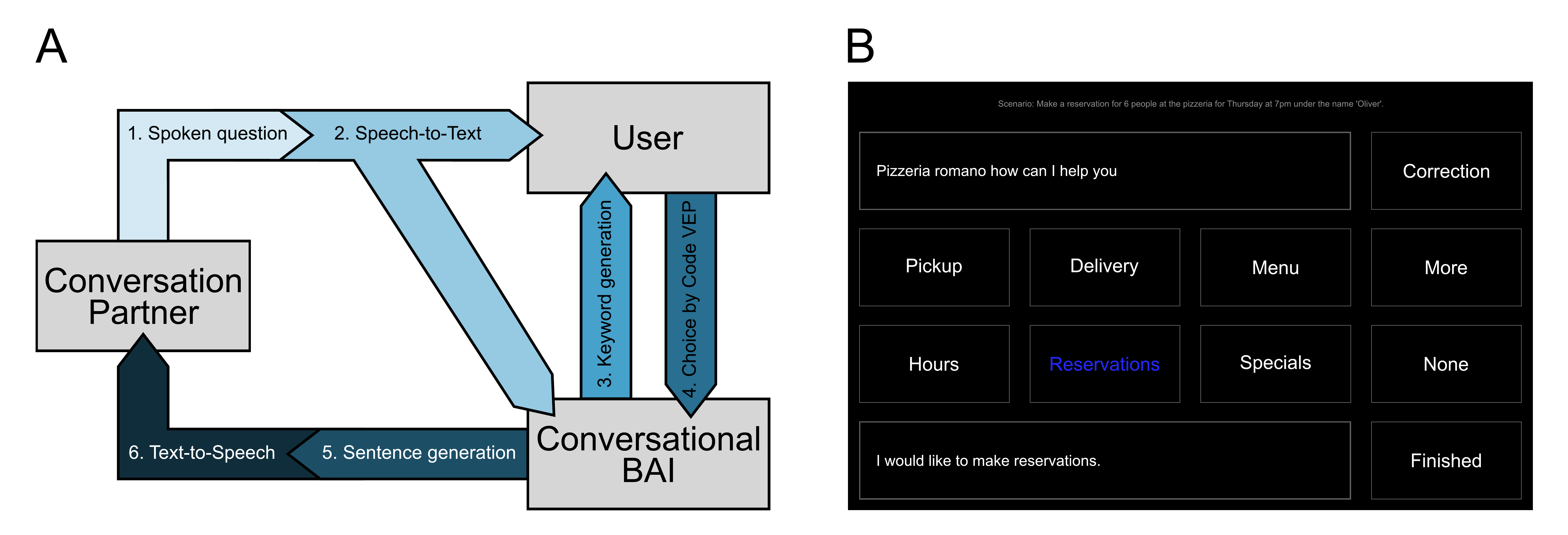}
    \caption{Structure \textbf{(A)} and screenshot \textbf{(B)} of EEGChat, the first Conversational BAI. One question-answer exchange occurs as follows: The conversation partner asks a question, which is automatically transcribed and supplied to both the user and BAI (displayed in the top left box on the screen). The Conversational BAI generates short possible answers and displays them to the user. Additionally, there are four special options for the user to correct mistakes ("Correction"), show more keywords ("More"), choose no keyword ("None"), or end the conversation ("Finished"). After reading the options, the user focuses on their chosen answer during a code-VEP stimulation phase. The choice is decoded from EEG and supplied to the BAI. The BAI then generates an appropriate answer to the question using the conversation history and chosen keyword. This full-sentence answer is displayed on the bottom left of the screen and played as a text-to-speech generated audio. The current scenario goal was displayed at the top of the screen during the experiment.}
    \label{fig:baiproto}
\end{figure}

\paragraph{Contextual input} When a conversation partner asks the BAI user a question, the first step is to input this contextual information into the system. We achieve this with automatic speech-to-text transcription, as shown in Figure~\ref{fig:baiproto}.A. The transcribed question is also displayed for the user on a screen, as shown in Figure~\ref{fig:baiproto}.B. If applicable, EEGChat retains the conversation history of previous question and answer pairs.

\paragraph{Cognitive probing} Upon receiving the transcribed question, the BAI must obtain information about the user's intention. Due to the difficulty of directly decoding user intentions from EEG data and the limitations of current conversational AIs, we chose an indirect cognitive probing method that allows the user to express a wide range of concepts. The conversational agent automatically generates several short answers (subsequently referred to as keywords) by querying the large language model GPT-3.5 \autocite{gpt3}. We developed a prompt to generate a variety of different answers that are relevant to the given question. Section~\ref{sec:methods_keywords} provides details of the keyword generation process.

The keywords are displayed to the user on the screen, as shown in Figure~\ref{fig:baiproto}.B. The number of presented keywords is adaptable. We chose a layout displaying six keyword options for the evaluation experiment (Section~\ref{sec:expdesign}). Additionally, the user is shown four special options to be able to correct mistakes ("Correction"), show more keywords ("More"), choose no keyword, and reply with "I am sorry, I cannot answer this question right now" ("None"), or end the conversation ("Finished").

To address questions requiring specific details like names and addresses, we integrated a knowledge base into EEGChat that enables users to provide personal information a-priori that they may want to disclose during a subsequent conversation. During keyword generation, we prompted the LLM to assign each question a category, e.g., NAME, NUMBER, or YESNO. If the category matches a knowledge base entry, the respective entries are displayed as choices instead of the generated keywords. In our experiment, the knowledge base included fictional names and addresses. In a real-world setting, users would fill these out in advance, understanding that this data might be shared with conversation partners.

\paragraph{Intention decoding} The next step is to infer from the user's neurophysiological data which of the ten options (six keywords and four special options) the user intends to choose. For this, we use a code-modulated visually evoked potential (code-VEP) paradigm \autocite{sutter92, bin2011high} with a decoding pipeline based on the work by \textcite{nagel2019plos}. Details of the decoding pipeline are described in Section~\ref{sec:methods_decoding}. After the user has had time to read the options, the boxes containing the stimuli start flickering in an on-off pattern with an individual time shift for every stimulus. The user focuses their attention on their choice, which elicits a corresponding pattern of VEPs in occipital EEG channels from which the attended keyword can be inferred. 

\paragraph{Action generation} 
After the BAI is aware of the user's intention, a full-sentence answer that matches the conversation history and the intention of the chosen keyword must be generated. We adapted the language model GPT-3.5 \autocite{gpt3} by fine-tuning it on a custom data set of conversations annotated with relevant keywords. This strongly improved the ability of the AI agent to identify which viewpoint it should adopt in the conversation as well as decrease the amount of additional wrong information included in the generated answers, compared to a non-tuned version. Refer to Section~\ref{sec:methods_full_sentence} for a detailed description of the data sets, fine-tuning approach, and model evaluation.

This final response is displayed on the screen as shown in Figure~\ref{fig:baiproto}.B and played as an audio output generated by a text-to-speech model. Upon hearing the BAI user's answer, the conversation partner can respond with another question to continue the conversation.

\subsection{Experimental design}
\label{sec:expdesign}

\begin{figure}[htb]
    \centering
    \includegraphics[width=\textwidth]{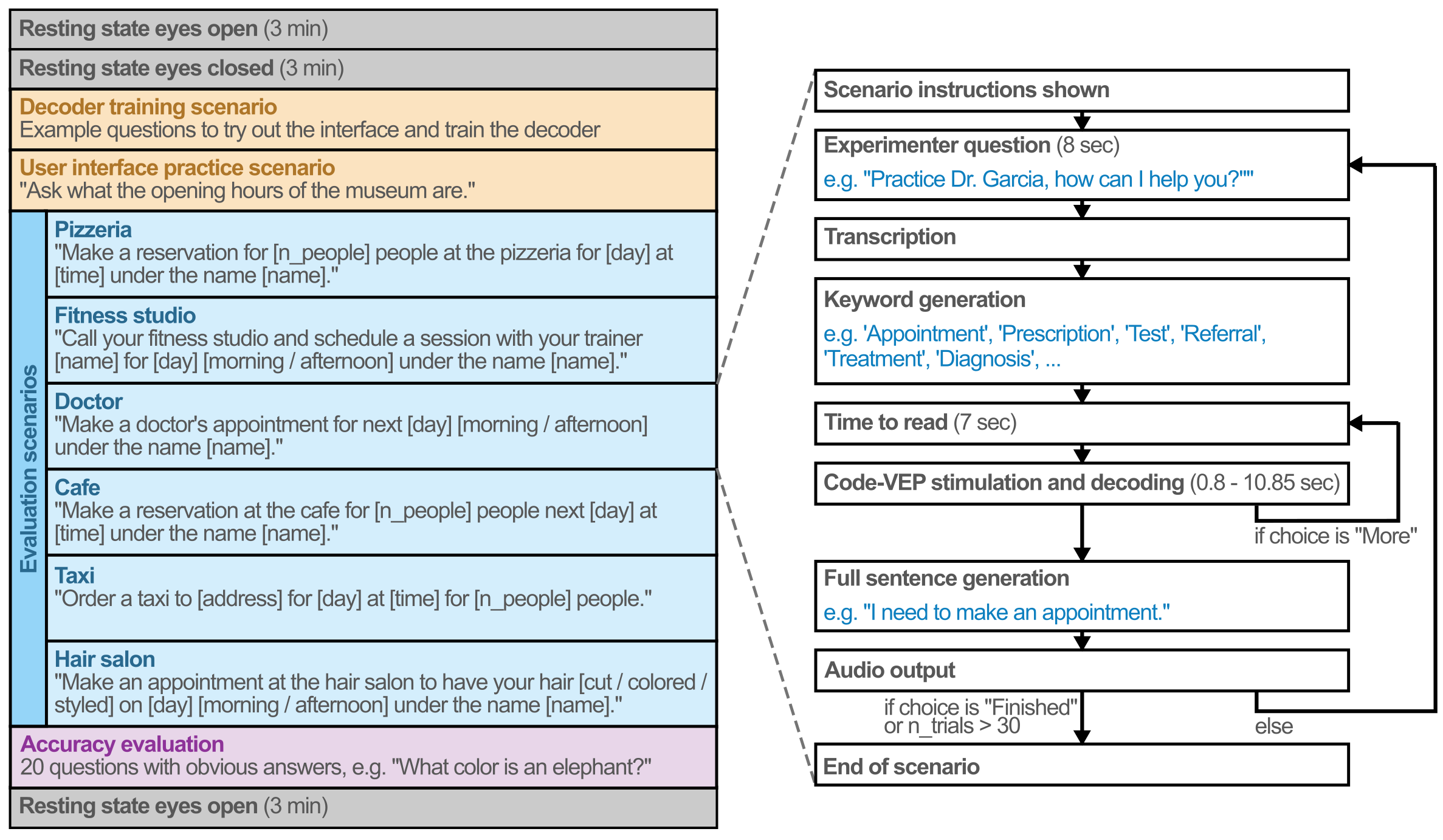}
     \caption{A detailed view of the experiment, consisting of a classifier training and UI familiarization phase, an evaluation scenario phase, and a classifier accuracy evaluation phase. The structure of each scenario is shown on the right.}
    \label{fig:experiment}
\end{figure}

We designed an empirical study to test the applicability of our system in realistic situations. The subject uses EEGChat in a simulated phone conversation to achieve a predetermined goal, such as reserving a table in a restaurant. We emphasize that EEGChat was not trained on any of the chosen scenarios to ensure that the reported success rate in achieving a conversational goal generalizes to novel, real-world scenarios. As shown in Figure~\ref{fig:experiment}, the experiment consists of three parts: The training phase, the conversational task evaluation phase, and the post-evaluation phase.

\paragraph{Classifier and user training} The training data for the code-VEP decoder is collected while the experimenter guides the subject interactively through the user interface. The subject is instructed in this part with tasks such as "Please select 'None' now" or simple questions such as "Hello, how are you doing today?". The participant focuses their attention on their choice during the maximal stimulation time of 10.85 seconds (217 frames). Subsequently, and only in this part of the experiment, the subject clicks on their selection to provide the true label for classifier training. 

The questions are semi-scripted and can be adapted to ensure the participant understands how to use the BAI and that the classifier has reached a good accuracy. Every participant made at least seven interactions with the system, therefore providing at least 1519 frames of training data.

After the classifier training, the subject completes a short practice scenario. Here, the subject witnesses how the system behaves during an evaluation scenario and tests if they can control it without the experimenter's guidance. The fully trained classifier is used for the first time in this scenario.

\paragraph{Conversational tasks} For each conversational task, the participant is instructed to complete realistic everyday scenarios using only the BAI (see Figure~\ref{fig:experiment} for the full list of scenarios). The subject receives a detailed goal description, whereas the conversation partner only knows the general topic of the task. For example, the conversation partner may be aware that the subject is calling a pizzeria, while the subject is also instructed to reserve a table for a specific number of people, time, and day. The experimenter speaks the first sentence of each scenario. They simulate picking up the phone and introduce themselves with a phrase such as "Pizzeria Romano, how can I help you?". Introducing the scenario in this way is necessary as it gives the BAI the relevant context of having called a pizzeria, something the subject, but not the system knows. To complete the scenario, the participant must successfully convey all relevant details to the experimenter during the conversation within at most 30 selections.\footnote{One selection does not necessarily mean one utterance due to the "More" option.} 

\paragraph{Accuracy evaluation \& questionnaire} After completing the conversational tasks, we assess the accuracy of the trained classifier. Subjects answer 20 simple, predetermined questions with obvious answers, such as "What color is an elephant?". Finally, the subjects are asked to complete a questionnaire asking about their experience using the BAI.

\subsection{Experimental results}
\label{sec:analysis}
Five healthy subjects (labeled S1 to S5) attempted the six conversation tasks shown in Figure \ref{fig:experiment}. Subjects S1-S4 accomplished the scenario goal in all conversations, while subject S5 failed in two out of six tasks due to misclassifications by the code-VEP decoder (S5 unintentionally selected \textit{Finished} too early in one task and selected an incorrect option in another task, which they were not able to correct. A video showcasing a representative conversation task, as well as all conversation transcripts, are provided in the supplementary material.

\begin{figure}[htb!]
    \centering
    \includegraphics[width=\textwidth]{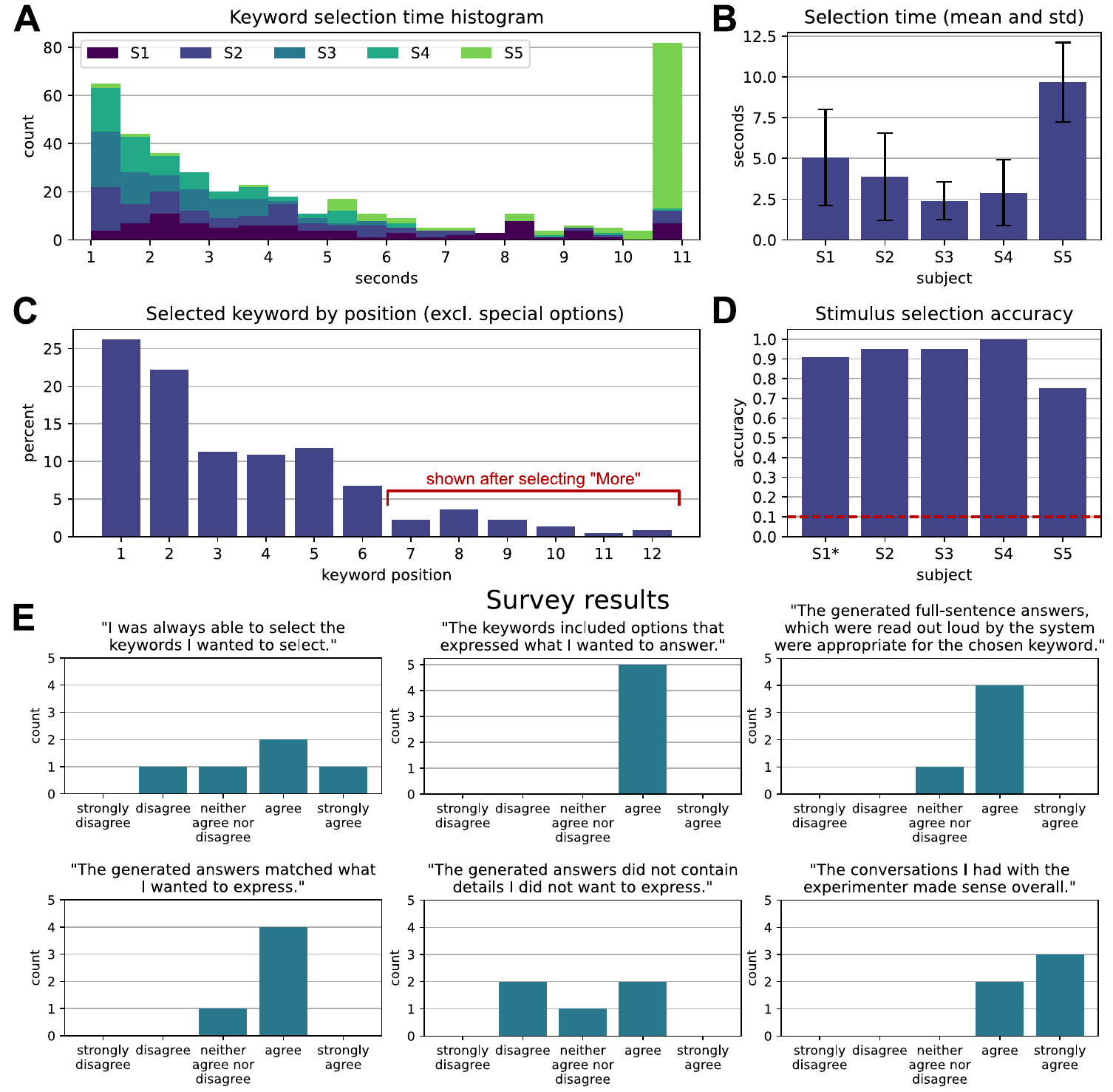}
    \caption{\textbf{(A)} Histogram of the time until a keyword was selected. Due to their low accuracy, S5 often reached the maximum time (10.85 s). \textbf{(B)} Mean and standard deviation of keyword selection time per subject. \textbf{(C)} Percentage of times a keyword at the given position was selected (excluding special options). Keywords 7 - 12 were only shown after selecting "More". \textbf{(D)} Stimulus selection accuracy during the accuracy evaluation stage, per subject. Accuracy was computed with 11 questions for S1 and with 20 questions for S2 - S5. Red line indicates chance level. \textbf{(E)} Results from the post-experiment participant survey.}
    \label{fig:bairesults}
\end{figure}

Successful tasks were completed in an average of 8.68 utterances, with a standard deviation of 3.43. As shown in Figure~\ref{fig:bairesults}.A--B, average selection times across subjects ranged from 2.23 seconds (S3) to 9.69 seconds (S5). A posthoc analysis revealed a preference for the first two generated keywords, which were selected $48\%$ of the time (Figure~\ref{fig:bairesults}.C). This demonstrates that LLMs, due to their tendency to suggest more likely tokens first \autocite{radford_language_nodate}, are well suited for fast and intuitive communication with BAIs. 

In the post-experiment survey, the results of which are summarized in Figure \ref{fig:bairesults}.E, all subjects reported understanding the goal and tasks of the experiment well. Subjects found the generated keywords relevant to the conversation and agreed that the keywords included options corresponding to their desired answers. Opinions on the full-sentence responses generated from the keyword varied, which could be attributed to the randomness of LLMs \autocite{radford_language_nodate}. The complete results of the survey are provided in the supplementary material.

\section{Discussion}

\label{sec:discussion}
Brain-Artificial Intelligence Interfaces establish a novel research direction in the area of neuroprostheses. By delegating significant parts of the planning and generation of actions to an AI agent while leaving high-level control with the user, complex actions can be taken with relatively little information extracted from neurophysiological data. In the case of EEGChat, the BAI approach enables users to engage in complex, open-topic conversations, all while utilizing EEG data and relying on relatively simple decoding paradigms. This is a significant advancement over earlier work in (invasive) BCI speech prostheses, which achieved a decoding performance that is sufficiently high for fluent conversation on small vocabularies only, limiting users' ability to engage in complex, open-topic conversations \cite{willett2023high}.

Furthermore, by replacing parts of the cognitive processing pipeline, people with severe cortical deficits who are not served by more traditional BCI architectures can benefit from this new type of assistive technology. Aphasia, the loss of the ability to comprehend or formulate language, for example, ranks among the health conditions most detrimental to an individual's health-related quality of life \cite{lam2010relationship}. The prospect of restoring such lost cognitive functions with the aid of AI substantially enlarges the target group that can benefit from this technology.

While advances in AI in general and in LLMs in particular have been revolutionary, LLMs are currently generic text generators. Their performance is likely to improve dramatically when fine-tuned with task-specific training data. The same applies to LLMs used in Conversational BAIs. Fine-tuning these models with user-specific conversational data, e.g., as obtained from past social media and email communications, is likely to dramatically improve their performance and replicate users' individual communication styles. At the same time, novel recording techniques for brain imaging and improved decoding pipelines may eventually eliminate the need for explicit cognitive probing, allowing for implicit, intuitive control of Conversational BAIs. However, the potential of BAIs extends beyond the replacement of lost cognitive functions. Leveraging AI agents to extend human capabilities may be an interesting future application area for BAIs. As research of AI agents in established and emerging fields progresses, it will likely enhance the efficacy and potential of BAIs, transforming how we communicate, perceive, and interact with technology.

At the same time, the highly innovative nature and possible impact of BAIs warrant a careful consideration of ethical implications. For instance, it must be ensured that users control the information they share via a BAI, similar to natural speech, and can even lie if they wish to do so. Further, integrating AIs into neuroprostheses renders them vulnerable to adversarial attacks, e.g., based on prompt-injection \cite{wei2023jailbroken, perez2022ignore, greshake2023not}. Finally, users relying on such technology for communication might want to take legally binding actions. Determining the responsibility for actions conducted with a BAI is critical. 

Despite these open questions, BAIs hold immense potential for enhancing the quality of life for numerous individuals. The work presented here paves the way for new avenues of research in developing advanced assistive devices, benefiting a large, underserved population.

\section{Methods}
\label{sec:methods}
This section offers a comprehensive overview of the components employed in EEGChat. In Section~\ref{sec:methods_audio}, we first outline our selection rationale in choosing the models utilized in audio processing of the speech in- and output. Subsequently, we delve into the methodologies employed for answer generations, encompassing a keyword generator for providing answer options during cognitive probing (Section~\ref{sec:methods_keywords}) and a full-sentence generator (Section~\ref{sec:methods_full_sentence}) that expands the selected keyword to a full answer sentence. Finally, in Section~\ref{sec:methods_decoding}, we discuss the rationale behind our adoption of code-VEP as the decoding pipeline and the tailored modification applied to align the decoder with our paradigm.

\subsection{Audio processing}
\label{sec:methods_audio}
Sentences spoken by conversation partners must the transcribed to text to interface with the large-language-models (LLMs). To achieve this, each trial begins with an eight-second window, during which the conversation partner speaks a question into the microphone. This recording, sampled at 16 kHz, undergoes transcription using the Google Cloud Speech API via the SpeechRecognition Python library \autocite{pythonspeechrec}. Should the transcription fail, EEGChat asks the conversation partner to repeat their question. While various transcription models exist, we opted for the Google Cloud Speech API for its simplicity and satisfactory performance.

Conversely, LLMs exclusively produce textual output. To convert this text into audible responses, we employed the Google Cloud Text-to-Speech API. This API, which given text returns an audio file with the text spoken by a synthetic voice, was chosen for its simplicity and effectiveness. While the current model suffices, exploring a customized version mimicking the sound of the user's voice remains an avenue for improvement.

\subsection{Text generation}
In the subsequent sections, we explain the details of the conversational agent at the core of EEGChat. Our system leverages transformer models, a class of neural networks first proposed by \textcite{vaswani2017attention}. They rely on the attention mechanism without any convolutional or recurrent units in the neural network. Their self-attention mechanism aggregates information from the input sequence, with the ability to handle long-term dependencies with no information loss even for the most distant parts in the input sequence \autocite{xu2021human}. 

The core system components employ the state-of-the-art language models available at the time of their creation each. For the keyword generation, OpenAI's GPT-3.5 \autocite{gpt3} is used, whereas, due to its earlier development, the full-sentence generation utilizes OpenAI's GPT-3 Davinci. Both models underwent testing with both modification techniques available, fine-tuning, and prompt-engineering. Ultimately, we adopted a prompt-engineered solution in the keyword generation and a custom fine-tuned version to produce the full-sentence answers, which we describe below. 

\subsubsection{Keywords}
\label{sec:methods_keywords}
\paragraph{Keyword generation}

The keyword options presented to the user are generated by a prompt-engineered (purposeful and systematic prompt designed) GPT-3.5 model, the state-of-the-art language model at the time. We found that for our use case, a balance between minimizing model randomness (lower temperature) and allowing for creativity in responses (higher temperature) by setting the model temperature to $0.5$ worked well. The engineered prompt comprises a paragraph describing the problem setting, conveying the task of generating keywords and the context -- assisting a speech-impaired person by providing responses to questions. To maintain consistency and guide the model, the prompt further includes examples of prompts and outputs. At the end, the prompt must be extended with a question and the desired number of generated keywords (twelve in our setting). For each query, the LLM outputs the specified number of keywords, as well as a category for the generated keywords, such as NAME, NUMBER, or YESNO, which is utilized in the knowledge base. The exact formulation of the prompt is included in the supplementary material. 

\paragraph{Knowledge base}
To give the user access to personal information, such as important addresses (e.g., home address or their doctor's address), EEGChat incorporates a knowledge base. The knowledge base is a dictionary mapping category strings to relevant options for that category. If the category returned by the keyword generation matches an entry in the knowledge base, the respective entries will be used as keyword options instead of the ones generated by the language model. During the experiment, the knowledge base contained entries for the categories NAME and ADDRESS, with lists of names and addresses vital for scenario completion during the experiment.

\paragraph{Special keyword options}
In addition to six keywords, EEGChat always displays four other options on the right side of the screen (see Figure~\ref{fig:baiproto}~(B)). These options assist the user in constructing an answer if none of the generated keywords present a suitable choice. The special options consist of:

\begin{itemize}
    \item \textbf{Correction:} EEGChat answers, "I am sorry, I misspoke earlier," giving the conversation partner the chance to repeat the previous question.
    \item \textbf{More / Previous:} Replace the six keywords with another 6 options / Show the previous six keywords
    \item \textbf{None:} The question cannot be answered with any of the twelve keywords (six original and six after clicking 'More'). EEGChat answers, "I am sorry, I cannot answer this question right now."
    \item \textbf{Finished:} The user has achieved their goal and wants to end the conversation. EEGChat answers, "Thank you, goodbye." After 'Finished' is selected, the scenario ends, and a new scenario begins.
\end{itemize}

\subsubsection{Full sentence response generation} 
\label{sec:methods_full_sentence}
To generate coherent and contextually relevant sentences from keywords, we employed GPT-3 Davinci, the latest GPT model available during the development of this component. GPT-3 is not inherently designed for generating sentences from keywords, necessitating additional steps to make the model behave in the desired way. These steps involve either a purposeful and systematic prompt design (prompt engineering) or creating a data set representative of the problem, which is then used to adjust the underlying model's weights (fine-tuning). Moreover, when these sentences are employed to speak for a user (instead of with them), it is crucial to establish strict criteria for the generated content. For instance, generating sentences with added information may lead to the user expressing opinions they do not agree with, potentially resulting in uncomfortable situations. 

In the following paragraphs, we delve into the specifics of prompt engineering and the fine-tuning procedures. We created five systems, employing different iterations of both techniques. To determine the optimal model, we established an evaluation process where recruited individuals assess response quality and identify model errors.

\paragraph{Training data for fine-tuning}
To construct the data set used for fine-tuning, we utilized the corpus summarized in the GitHub repository at \url{https://github.com/zeloru/small-english-smalltalk-corpus} as a foundation. This corpus comprises multiple English small-talk conversations, primarily transcriptions of English lesson conversation examples. This makes the corpus ideal for our use case, as most of the sentences are dialogues and, as the sentences are aimed at English learners, they tend to be rather simple, reducing complexity.

Before constructing any samples, we manually filtered the corpus, removing irrelevant and too short conversations, resulting in approximately 890 usable conversations, each comprising question-answer (QA) pairs between two individuals. These conversation were then used to construct three distinct data sets: FT-XL, a large data set using all 890 conversations; FT-CR, a more creative data set using 520 of the conversations, made more creative by providing fewer keywords (further discussed later); and FT-HQ, a high-quality data set, additionally edited by humans, comprising of 330 conversations.

The QA pairs from these conversations served as the foundation for our data set samples. Each sample of the data set includes all information available to the system during a potential deployment: a question posed by the conversation partner, one or more keywords for answer generation, an optional dialogue history consisting of past QA pairs, and, as the targeted output, the corresponding answer sentence. To prevent the training of overly lengthy responses, we introduced a condition discarding samples with answer sentences exceeding 110 characters. Further, we focused on shorter conversations by distributing the conversation history in the data set as follows: no history - 50\%, one QA pair of history - 35\%, two QA pairs of history - 8\%, three QA pairs of history - 7\%.

Keywords, as they are not part of the corpus, needed to be generated additionally. We have done so by randomly selecting one of the automated keyword extraction technologies -- YAKE \autocite{CAMPOS2020257}, RAKE \autocite{RAKE}, and KeyBERT \autocite{KeyBERT}. YAKE was additionally used with a setting to extract a few most relevant keywords (n=1, dedupLim=0.9) and another to extract a detailed list of keywords (n=5, dedupLim=0). If one of the methods was unable to extract any keywords, another was picked at random. If none of the methods were able to extract a keyword, the sample was rejected. Instead of the above described random picking, FT-CR always chose the result of the generator with the smallest number of keywords. To increase clarity and accuracy, contractions, such as \textit{I'll} or \textit{won't}, were expanded to their full forms \textit{I will} and \textit{will not}. We also addressed negations missing in the keywords by additionally adding them. After these steps, FT-HQ additionally enhances data set quality through manual review and adjustment conducted by three students, involving a total of 30 hours of effort.

Following data set creation, we saved them as JSON files to be used in the fine-tuning process later. These files comprise two fields: \textit{prompt}, containing conversation history, question, and keywords, and \textit{completion}, containing the answer sentence. Each question and answer in the history within the \textit{prompt} were delineated with \verb|Question:| and \verb|Answer:|, respectively, followed by a new line. After the history, the current question was provided after \verb|Question:|, followed by the chosen keywords after \verb|Keyword:|. Lastly, the \textit{prompt}, concluded with another \verb|Answer:| followed by \verb|\n\n###\n\n|, indicating to the model to generate an answer. The \textit{completion} only contained the answer sentence followed by an \verb|END| token, serving as the stop sequence in model deployment, ending generation when the model prints it. 

\paragraph{Fine-tuning}
The three above described data sets have been used to create GPT-3 Davinci \autocite{gpt3} fine-tunes. This model was chosen due to its status as the largest and best-performing model available at the time of development. The process of fine-tuning a GPT model involves adapting the pre-trained original model to a specific task. OpenAI provides flexibility in setting various hyperparameters during the fine-tuning process, including epochs, learning rate multiplier, and batch size. Through extensive experimentation with multiple settings, we decided on the following configuration for our use case: epochs - 1, learning rate multiplier - 0.05, and batch size - 2. This choice was made after careful consideration of the model's performance and the specific requirements of our project.

\paragraph{Evaluation}
We evaluated five distinct model versions, with three of them being trained on the data sets FT-XL, FT-CT, and FT-HQ, discussed previously. Additionally, we constructed two versions using prompt engineering, denoted as STOCK-1 and STOCK-2 (the exact prompts are included in the supplementary material). To assess the performance of these models, we designed an online experiment in which participants (recruited via Amazon Mechanical Turk and other platforms) rated the model outputs. 

The experiment was built around a web-based interface in which participants were tasked with evaluating responses generated by the different models. Each evaluation session began by introducing the participant to a predefined topic and presenting them with an initial question. For example: Topic - \textit{Somebody asks you about hobbies, talk with her about your hobbies} and Question - \textit{Hi! Let's chat about hobbies, what is yours?}. Participants were then asked to provide both a detailed, natural language response and a response composed solely of keywords. Subsequently, the evaluated system utilized the keywords and conversation history to generate a response to the question.

Participants were then tasked to evaluate how well the generated answer expressed their intentions on a scale of \textit{Wrong} (1), \textit{Bad} (2), \textit{Okay} (3), \textit{Good} (4), and \textit{Perfect} (5). If any option other than \textit{Perfect} was selected, participants were further prompted to specify what was unsatisfactory with the answer. Six options for unsatisfactory responses were provided: \textit{Slightly different}, \textit{Too brief/short}, \textit{Missed details}, \textit{Added details}, \textit{Wrong information}, and \textit{Totally different}. This evaluation process was repeated with three questions per presented scenario until a total of ten questions and evaluation pairs were acquired. After cleaning the data\footnote{Invalid responses could be generated on Amazon Mechanical Turk when individuals attempt to exploit the system for rewards without engaging in the experiment genuinely}, we retained 2455 valid answer and evaluation pairs.

\paragraph{Model performance}

\begin{figure}[htb!]
    \centering
    \includegraphics[width=\textwidth]{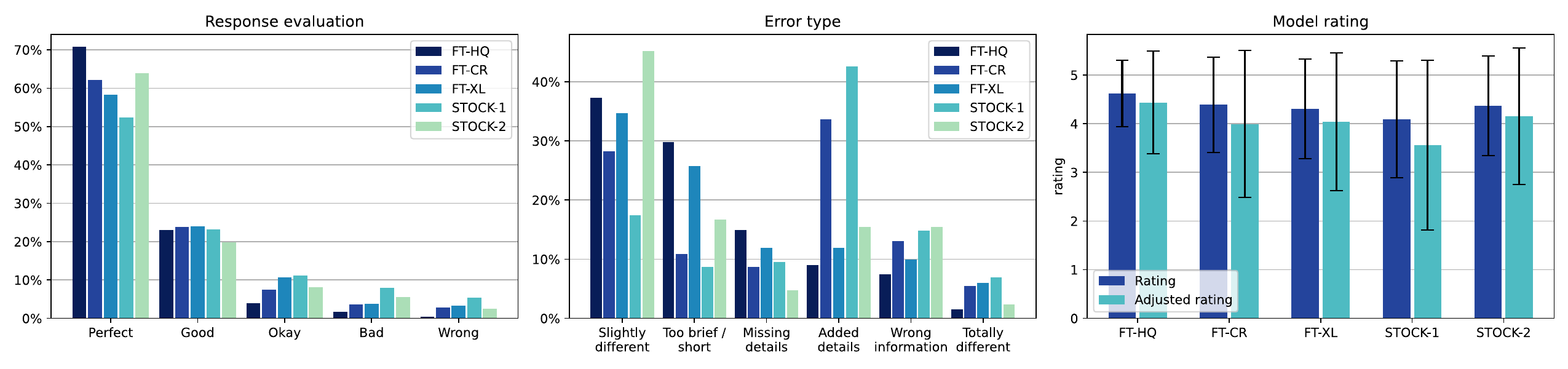}
    \caption{From left to right: Distribution of the performance metrics evaluated by human experiment participants, distribution of the mistakes causing a model to have a lower-than-perfect score, and results of the performance analysis of the different models.
    The rating was computed from the answer performance (perfect (5) to wrong (1)). For the adjusted rating, the original rating is multiplied by a factor depending on the mistake made (slightly different - 1, too brief/short - 0.9, missed details - $\frac{2}{3}$, added details - 0.5, wrong information - 0.25 and totally different 0.1) }
    \label{fig:dialokey}
\end{figure}

The results of the experiment are presented in Figure~\ref{fig:dialokey}. Initially, the models may appear to perform similarly, as they all exhibit a gradation from \textit{Perfect} (5) to \textit{Wrong} (1). However, it is essential to note the low proportion of answers generated by FT-HQ falling within \textit{Okay}, \textit{Bad}, and \textit{Wrong} categories. Analyzing the average rating allows us to quantify this and conclude that FT-HQ achieved the highest average rating at 4.6217, with the lowest standard deviation of 0.6870.

To assess the statistical significance of our data, we conducted an ANOVA analysis. ANOVA helps us determine whether the means of the different models are likely to have been generated by different distributions or the same distribution. The p-value obtained was 7.12e-07, significantly below the alpha level of 0.05. Thus, we can confidently reject the null hypothesis that the means of our data were generated by the same distribution.

For answers rated worse than \textit{Perfect}, subjects were asked to categorize the types of mistakes made by the model. The results of this are shown in Figure~\ref{fig:dialokey}. Notably, FT-CR, designed with creativity in mind, makes the most mistakes by adding details. Further, STOCK-2, a model designed using prompt engineering, exhibits the highest proportion of \textit{Slightly different} mistakes, but also a substantial number of \textit{Wrong information} and \textit{Added details} mistakes. The FT-HQ and FT-XL models demonstrate a gradation from \textit{Slightly different} to \textit{Totally different}.

Recognizing that not all mistakes carry the same weight, we adjusted the rating of the different mistakes with the following factors: \textit{Slightly different} - 1, \textit{Too brief/short} - 0.9, \textit{Missed details} - $\frac{2}{3}$, \textit{Added details} - 0.5, \textit{Wrong information} - 0.25 and \textit{Totally different} 0.1. We then compute new ratings for the answers by multiplying the rating score (\textit{Perfect} to \textit{Wrong}) with the adjustment factor of the mistakes. Thus, for example, if a rating of \textit{Good} was given and a mistake of \textit{Added details} was selected, the new score for this answer would be $4 \cdot 0.5 = 2$. This adjustment is necessary, as different mistakes have varying impacts. For instance, if unnecessary information to the answer is added, the user might express something they did not wish to, causing an uncomfortable situation. The resulting adjusted rating scores can be found in Figure~\ref{fig:dialokey}.

With the adjusted ratings, we can observe that the average score of FT-HQ is better than that of the other models. Based on this analysis, we have chosen to use FT-HQ as the full sentence generator in the Conversational BAI experiment in this paper.

\subsection{Decoding intentions from EEG}
\label{sec:methods_decoding}
In this section, we describe the methods used to decode the user's intentions. First, we present the details of electroencephalography (EEG), then we explain the stimulus mechanism of code-modulated visually evoked potentials (code-VEPs) and why we chose it, and finally, we detail the machine-learning pipeline used to infer the chosen stimulus, consisting of a combination of a CNN and sequence matching approach.

\subsubsection{EEG recording}
Electroencephalography (EEG) is one of the most commonly used neurophysiological recording techniques for brain-computer interfaces due to its non-invasiveness, high temporal resolution, and good spatial coverage \autocite{abiri2019review}. It measures electrical potentials via electrodes on the scalp. During our experiments, participants' EEG was recorded with the channels POz, PO3, PO4, Oz, O9, and O10 at a sampling rate of 1000 Hz using a Brain Products EasyCap and Bittium NeurOne Tesla amplifiers. The electrode locations were selected due to their proximity to the visual cortex and following an analysis of the contribution to the prediction performance in a pilot subject for the code-VEP paradigm.

\begin{figure}[tb!]
    \centering
    \includegraphics[width=\textwidth]{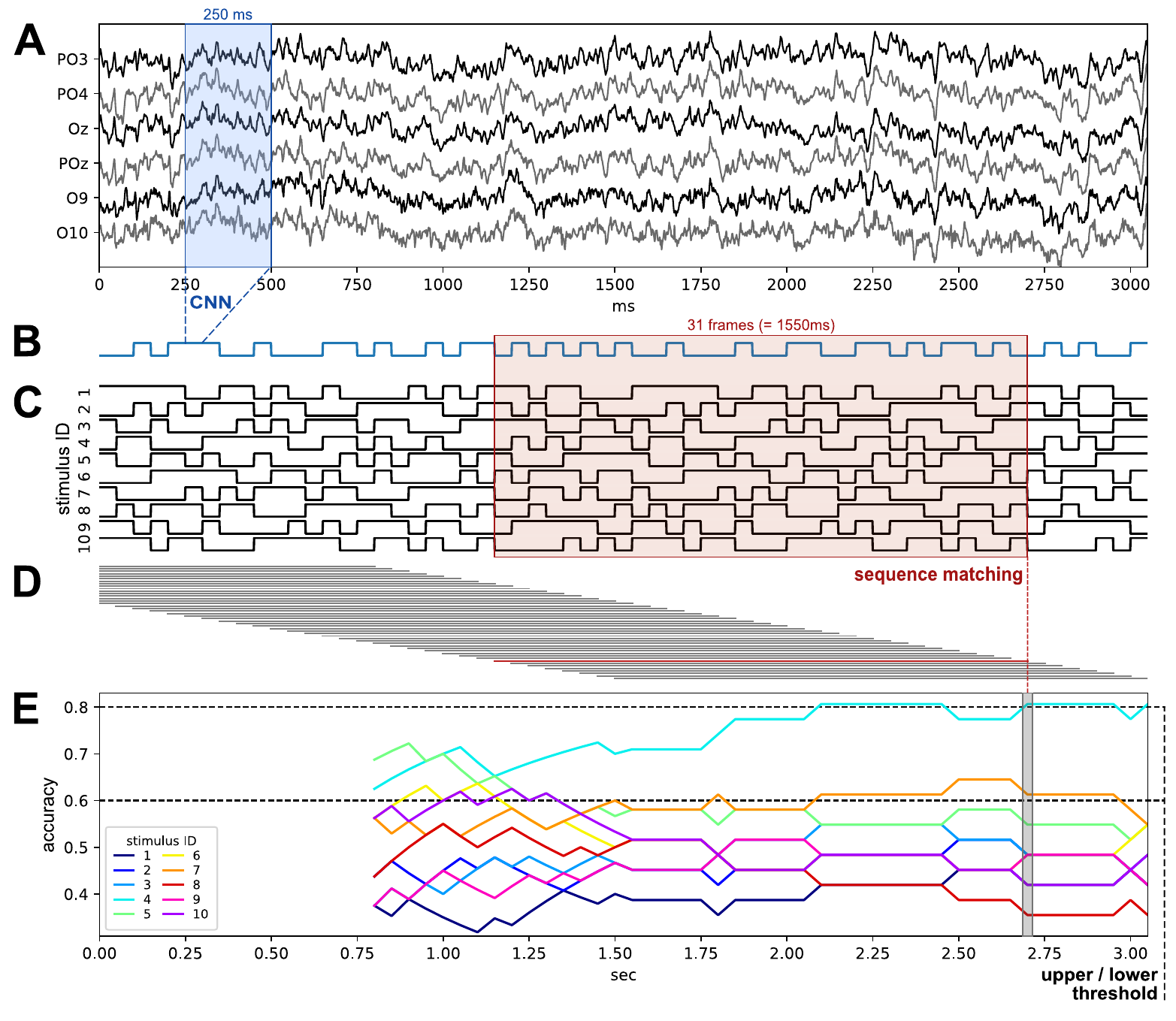}
    \caption{Process of decoding the attended stimulus from EEG data. During the stimulus presentation phase, each stimulus flashes in an individual on-off pattern \textbf{(C)}, where each frame is shown for 50 ms. The 250ms window of EEG data recorded during and after each frame presentation is fed into a CNN \textbf{(A)}. The CNN makes one prediction (on vs. off) per stimulus frame \textbf{(B)}. The sequence of predictions is compared to the stimulus patterns \textbf{(B, C)} on a rolling window basis \textbf{(D)} by computing the prediction accuracy \textbf{(E)}. If the accuracy of one stimulus reaches at least 0.8, while all others remain below 0.6, the respective stimulus is returned by the decoder.}
    \label{fig:decoding}
\end{figure}

\subsubsection{Stimulus paradigm (code-VEPs)}
The choice of stimulus paradigm was driven by several factors. Firstly, we needed a system with a short classification time, allowing the subject to quickly and easily choose the desired keywords. To be able to dynamically extend our system to arbitrary user interfaces, scalability to multiple classes was also considered. The chosen paradigm should also have a high classification accuracy and as short a training time as possible. Further, to ease setup, a system that requires fewer electrodes is preferred.

These criteria are all met with code-VEPs, which we have chosen as the paradigm for our experiment. Code-VEPs have shown a high accuracy with only one electrode in use \autocite{brainsci8070130}. However, better results are achieved if more electrodes are used \autocite{martinez-cagigal_braincomputer_2021}. A similar approach, steady state visually evoked potentials (SSVEPs), which flash the stimuli at constant frequencies, have a similar performance as code-VEPs \autocite{martinez-cagigal_braincomputer_2021}, but they, in general, do not scale well to multiple stimuli, as they are coupled to the monitor refresh rate. Meanwhile, code-VEPs can be easily extended to an arbitrary amount of stimuli by selecting codes with an autocorrelation of zero. 

The physiological mechanism utilized in code-VEPs is the brain's reaction to flashing visual stimuli. These visually evoked potentials (VEPs) are generally most strongly observed in the occipital lobe, where the brain's visual cortex is located. \textcite{sutter92} first proposed using this effect for BCIs, but only much later did the idea of code-modulated VEPs gain attention across the field \autocite{bin2011high, martinez2021}. During a code-VEP paradigm, a subject is presented with several stimuli, each flashing at a different on-off pattern. When the subject is focusing their attention on one of the stimuli, a corresponding pattern of evoked potentials can be observed in their EEG. 

During our experiment, the stimuli (boxes with keywords) were flashed in a time-shifted maximum length sequence (m-sequence) pattern of length $m=31$. Maximum length sequences are binary pseudorandom sequences with the advantageous property of having a circular autocorrelation of zero for every shift ${n \in \{1,...,m-1\}}$ \cite{bin2011high}. The stimulus pattern was flashed at a rate of 20Hz and 50\% amplitude, which has been shown to reduce strain on the eyes with only minimal loss in decoding accuracy \cite{ladouce2022improving, Castillos2023}. During each trial, the pattern was repeated a maximum of seven times, equaling at most 217 frames, or 10.85 seconds.

\subsubsection{Real-time keyword choice decoding}
During the flashing stimulus presentation, the choice of the user is decoded in real time using a two-step process. For every stimulus frame, a convolutional neural network (CNN) predicts whether the attended stimulus was on or off. This predicted sequence is matched continuously against the true stimulation patterns of all ten stimuli to determine which stimulus was attended to by the BAI user. 

The CNN used to decode the sequence of on-off patterns is EEG2Code \cite{nagel2019plos}, consisting of three convolution / max-pooling blocks and two dense layers. In contrast to to EEG2Code, we apply a window shift of 50 ms (flashing at 20 Hz), whereas they present a stimulus every 16.67 ms (flashing at 60 Hz). We predict each stimulus frame state from 250 ms (250 samples due to recording at 1000 Hz) windows of unfiltered EEG during and after the respective stimulus frame (50 ms + 200 ms), as shown in Figure~\ref{fig:decoding}.A-B. The chosen window length is in accordance with the original application of EEG2Code, where it is argued that a VEP signal usually lasts about 250 ms.  

After 16 frames have been presented (800 ms) and the full window of EEG of the \nth{16} frame has been recorded (+200 ms), the predicted sequence is continuously compared to the ten stimulation sequences by computing the prediction accuracy. The accuracy is computed for a shifted window of the most recent 31 frames at most, as shown in Figure~\ref{fig:decoding}.B-D. If the accuracy reaches at least 0.8 for one of the stimulus patterns, while the accuracy of all other sequences remains below 0.6, the decoder returns the respective stimulus ID, and stimulation stops (see Figure~\ref{fig:decoding}.E). The tenth stimulus ("Finished") is an exception since it ends the conversation. Erroneously choosing it can, therefore, not be recovered from through corrections. For it to be selected, the thresholds have to be crossed for a total of ten stimulus frames. If the thresholds are not met after seven repetitions of the stimulus patterns (10.85 s), the stimulus with the highest correlation to the predicted sequence is selected.

\section{Code and data availability}
The code of EEGChat and for GPT fine-tuning, as well as all experiment data, is available at \url{https://github.com/AKMeunier/EEGChat}.

\section {Ethical compliance}
The study protocol was evaluated and approved by the Ethics Committee of the University of Vienna prior to the start of the study. Informed consent was obtained from all participants.

\section{Acknowledgments}
We would like to thank Kalou Cabrera Castillos, Frédéric Dehais, Tryggvi Edwald, Peter Fugger, Philipp Raggam, and Nike Schödl for fruitful discussions, helpful feedback, and technical support.

\printbibliography

@inproceedings{madotto2020plug,
    title = "Plug-and-Play Conversational Models",
    author = "Madotto, Andrea and Ishii, Etsuko and Lin, Zhaojiang and Dathathri, Sumanth and Fung, Pascale",
    editor = "Cohn, Trevor  and
      He, Yulan  and
      Liu, Yang",
    booktitle = "Findings of the Association for Computational Linguistics: EMNLP 2020",
    month = nov,
    year = "2020",
    publisher = "Association for Computational Linguistics",
    pages = "2422--2433",
}

@article{dialogpt,
  title={Dialogpt: Large-scale generative pre-training for conversational response generation},
  author={Zhang, Yizhe and Sun, Siqi and Galley, Michel and Chen, Yen-Chun and Brockett, Chris and Gao, Xiang and Gao, Jianfeng and Liu, Jingjing and Dolan, Bill},
  journal={arXiv preprint arXiv:1911.00536},
  year={2019}
}

@article{vaswani2017attention,
  title={Attention is all you need},
  author={Vaswani, Ashish and Shazeer, Noam and Parmar, Niki and Uszkoreit, Jakob and Jones, Llion and Gomez, Aidan N and Kaiser, {\L}ukasz and Polosukhin, Illia},
  journal={Advances in neural information processing systems},
  volume={30},
  year={2017}
}

@book{Levelt89Speaking,
    year = {1989},
    title = {Speaking: From Intention to Articulation},
    language = {eng},
    address = {Cambridge, Massachusetts},
    author = {Willem J. M. Levelt},
    publisher = {A Bradford Book},
}

@book{Mayer2006Automaticity,
    year={2006},
    title={Automaticity and Control in Language Processing},
    author={A. Meyer and L. Wheeldon and A. Krott},
    publisher={Psychology Press},
    address={London}
}

@article{gpt3,
  title={Language models are few-shot learners},
  author = {Brown, Tom B. and Mann, Benjamin and Ryder, Nick and Subbiah, Melanie and Kaplan, Jared and Dhariwal, Prafulla and Neelakantan, Arvind and Shyam, Pranav and Sastry, Girish and Askell, Amanda and Agarwal, Sandhini and Herbert-Voss, Ariel and Krueger, Gretchen and Henighan, Tom and Child, Rewon and Ramesh, Aditya and Ziegler, Daniel M. and Wu, Jeffrey and Winter, Clemens and Hesse, Christopher and Chen, Mark and Sigler, Eric and Litwin, Mateusz and Gray, Scott and Chess, Benjamin and Clark, Jack and Berner, Christopher and McCandlish, Sam and Radford, Alec and Sutskever, Ilya and Amodei, Dario},
  journal={Advances in neural information processing systems},
  volume={33},
  pages={1877--1901},
  year={2020}
}

@article{angrick2019,
  title={Speech synthesis from ECoG using densely connected 3D convolutional neural networks},
  author={Angrick, Miguel and Herff, Christian and Mugler, Emily and Tate, Matthew C and Slutzky, Marc W and Krusienski, Dean J and Schultz, Tanja},
  journal={Journal of neural engineering},
  volume={16},
  number={3},
  pages={036019},
  year={2019},
  publisher={IOP Publishing}
}

@article{anumanchipalli2019,
  title={Speech synthesis from neural decoding of spoken sentences},
  author={Anumanchipalli, Gopala K and Chartier, Josh and Chang, Edward F},
  journal={Nature},
  volume={568},
  number={7753},
  pages={493--498},
  year={2019},
  publisher={Nature Publishing Group}
}

@article{proix2022imagined,
  title={Imagined speech can be decoded from low-and cross-frequency intracranial EEG features},
  author={Proix, Timoth{\'e}e and Delgado Saa, Jaime and Christen, Andy and Martin, Stephanie and Pasley, Brian N and Knight, Robert T and Tian, Xing and Poeppel, David and Doyle, Werner K and Devinsky, Orrin and others},
  journal={Nature communications},
  volume={13},
  number={1},
  pages={1--14},
  year={2022},
  publisher={Nature Publishing Group}
}

@article{wolpaw2000brain,
  title={Brain-computer interface technology: A review of the first international meeting},
  author={Wolpaw, Jonathan R and Birbaumer, Niels and Heetderks, William J and McFarland, Dennis J and Peckham, P Hunter and Schalk, Gerwin and Donchin, Emanuel and Quatrano, Louis A and Robinson, Charles J and Vaughan, Theresa M and others},
  journal={IEEE Transactions on Rehabilitation Engineering},
  volume={8},
  number={2},
  pages={164--173},
  year={2000}
}

@article{birbaumer2000thought,
  title={The thought translation device (TTD) for completely paralyzed patients},
  author={Birbaumer, Niels and Kubler, Andrea and Ghanayim, Nimr and Hinterberger, Thilo and Perelmouter, Jouri and Kaiser, Jochen and Iversen, Iver and Kotchoubey, Boris and Neumann, Nicola and Flor, Herta},
  journal={IEEE Transactions on Rehabilitation Engineering},
  volume={8},
  number={2},
  pages={190--193},
  year={2000},
  publisher={IEEE}
}

@article{zander2011towards,
  title={Towards passive brain--computer interfaces: applying brain--computer interface technology to human--machine systems in general},
  author={Zander, Thorsten O and Kothe, Christian},
  journal={Journal of Neural Engineering},
  volume={8},
  number={2},
  pages={025005},
  year={2011},
  publisher={IOP Publishing}
}

@article{grosse2011using,
  title={Using brain--computer interfaces to induce neural plasticity and restore function},
  author={Grosse-Wentrup, Moritz and Mattia, Donatella and Oweiss, Karim},
  journal={Journal of Neural Engineering},
  volume={8},
  number={2},
  pages={025004},
  year={2011},
  publisher={IOP Publishing}
}

@article{gomez2011closing,
  title={Closing the sensorimotor loop: haptic feedback facilitates decoding of motor imagery},
  author={Gomez-Rodriguez, Manuel and Peters, Jan and Hill, J and Sch{\"o}lkopf, Bernhard and Gharabaghi, Alireza and Grosse-Wentrup, Moritz},
  journal={Journal of Neural Engineering},
  volume={8},
  number={3},
  pages={036005},
  year={2011},
  publisher={IOP Publishing}
}

@article{lawrence2001estimates,
  title={Estimates of the prevalence of acute stroke impairments and disability in a multiethnic population},
  author={Lawrence, Enas S and Coshall, Catherine and Dundas, Ruth and Stewart, Judy and Rudd, Anthony G and Howard, Robin and Wolfe, Charles DA},
  journal={Stroke},
  volume={32},
  number={6},
  pages={1279--1284},
  year={2001},
  publisher={Am Heart Assoc}
}

@article{dickey2010incidence,
  title={Incidence and profile of inpatient stroke-induced aphasia in Ontario, Canada},
  author={Dickey, Laura and Kagan, Aura and Lindsay, M Patrice and Fang, Jiming and Rowland, Alexandra and Black, Sandra},
  journal={Archives of physical medicine and rehabilitation},
  volume={91},
  number={2},
  pages={196--202},
  year={2010},
  publisher={Elsevier}
}

@article{russakovsky2015imagenet,
  title={Imagenet large scale visual recognition challenge},
  author={Russakovsky, Olga and Deng, Jia and Su, Hao and Krause, Jonathan and Satheesh, Sanjeev and Ma, Sean and Huang, Zhiheng and Karpathy, Andrej and Khosla, Aditya and Bernstein, Michael and others},
  journal={International journal of computer vision},
  volume={115},
  number={3},
  pages={211--252},
  year={2015},
  publisher={Springer}
}

@inproceedings{deng2013new,
  title={New types of deep neural network learning for speech recognition and related applications: An overview},
  author={Deng, Li and Hinton, Geoffrey and Kingsbury, Brian},
  booktitle={2013 IEEE international conference on acoustics, speech and signal processing},
  pages={8599--8603},
  year={2013},
  organization={IEEE}
}

@article{mnih2015human,
  title={Human-level control through deep reinforcement learning},
  author={Mnih, Volodymyr and Kavukcuoglu, Koray and Silver, David and Rusu, Andrei A and Veness, Joel and Bellemare, Marc G and Graves, Alex and Riedmiller, Martin and Fidjeland, Andreas K and Ostrovski, Georg and others},
  journal={Nature},
  volume={518},
  number={7540},
  pages={529--533},
  year={2015},
  publisher={Nature Publishing Group}
}

@article{meta2022human,
  title={Human-level play in the game of Diplomacy by combining language models with strategic reasoning},
  author={Meta Fundamental AI Research Diplomacy Team (FAIR) and Bakhtin, Anton and Brown, Noam and Dinan, Emily and Farina, Gabriele and Flaherty, Colin and Fried, Daniel and Goff, Andrew and Gray, Jonathan and Hu, Hengyuan and others},
  journal={Science},
  year={2022},
  publisher={American Association for the Advancement of Science}
}

@misc{pythonspeechrec,
    title = "Speech Recognition (Version 3.8)",
    author = "Zhang, A." ,
    year = "2017",
    note = "Software",
    url = "https://github.com/Uberi/speech_recognition"
}

@article{nagel2019plos,
    author = {Nagel, Sebastian AND Spüler, Martin},
    journal = {PLOS ONE},
    publisher = {Public Library of Science},
    title = {World’s fastest brain-computer interface: Combining EEG2Code with deep learning},
    year = {2019},
    month = {09},
    volume = {14},
    pages = {1-15},
    number = {9}
}

@article{sutter92,
title = {The brain response interface: communication through visually-induced electrical brain responses},
journal = {Journal of Microcomputer Applications},
volume = {15},
number = {1},
pages = {31-45},
year = {1992},
note = {Special Issue on Computers for Handicapped People},
author = {Erich E. Sutter}
}

@article{abiri2019review,
  title={A comprehensive review of EEG-based brain--computer interface paradigms},
  author={Abiri, Reza and Borhani, Soheil and Sellers, Eric W and Jiang, Yang and Zhao, Xiaopeng},
  journal={Journal of neural engineering},
  volume={16},
  number={1},
  pages={011001},
  year={2019},
  publisher={IOP Publishing}
}

@article{martinez2021,
  title={Brain--computer interfaces based on code-modulated visual evoked potentials (c-VEP): A literature review},
  author={Martínez-Cagigal, Víctor and Thielen, Jordy and Santamaria-Vazquez, Eduardo and P{\'e}rez-Velasco, Sergio and Desain, Peter and Hornero, Roberto},
  journal={Journal of Neural Engineering},
  volume={18},
  number={6},
  pages={061002},
  year={2021},
  publisher={IOP Publishing}
}

@article{willett2023high,
  title={A high-performance speech neuroprosthesis},
  author={Willett, Francis R and Kunz, Erin M and Fan, Chaofei and Avansino, Donald T and Wilson, Guy H and Choi, Eun Young and Kamdar, Foram and Glasser, Matthew F and Hochberg, Leigh R and Druckmann, Shaul and others},
  journal={Nature},
  pages={1--6},
  year={2023},
  publisher={Nature Publishing Group UK London}
}

@article{metzger2023high,
  title={A high-performance neuroprosthesis for speech decoding and avatar control},
  author={Metzger, Sean L and Littlejohn, Kaylo T and Silva, Alexander B and Moses, David A and Seaton, Margaret P and Wang, Ran and Dougherty, Maximilian E and Liu, Jessie R and Wu, Peter and Berger, Michael A and others},
  journal={Nature},
  pages={1--10},
  year={2023},
  publisher={Nature Publishing Group UK London}
}

@article {Castillos2023,
	author = {K. Cabrera Castillos and S. Ladouce and L. Darmet and F. Dehais},
	title = {Burst c-VEP Based BCI: Optimizing Stimulus Design for Enhanced Classification with Minimal Calibration Data and Improved User Experience},
	elocation-id = {2023.08.18.553900},
	year = {2023},
	publisher = {Cold Spring Harbor Laboratory},
	journal = {bioRxiv}
}

@article{ladouce2022improving,
  title={Improving user experience of SSVEP BCI through low amplitude depth and high frequency stimuli design},
  author={Ladouce, Simon and Darmet, Ludovic and Torre Tresols, Juan Jesus and Velut, S{\'e}bastien and Ferraro, Giuseppe and Dehais, Fr{\'e}d{\'e}ric},
  journal={Scientific Reports},
  volume={12},
  number={1},
  pages={8865},
  year={2022},
  publisher={Nature Publishing Group UK London}
}

@article{bin2011high,
  title={A high-speed BCI based on code modulation VEP},
  author={Bin, Guangyu and Gao, Xiaorong and Wang, Yijun and Li, Yun and Hong, Bo and Gao, Shangkai},
  journal={Journal of neural engineering},
  volume={8},
  number={2},
  pages={025015},
  year={2011},
  publisher={IOP Publishing}
}

@article{CAMPOS2020257,
title = {YAKE! Keyword extraction from single documents using multiple local features},
journal = {Information Sciences},
volume = {509},
pages = {257-289},
year = {2020},
issn = {0020-0255},
author = {Ricardo Campos and Vítor Mangaravite and Arian Pasquali and Alípio Jorge and Célia Nunes and Adam Jatowt},
}

@article{RAKE,
	year = 2019,
	month = {August},
	publisher = {Preprints},
	author = {Prafull Sharma and Yingbo Li},
	title = {Self-Supervised Contextual Keyword and Keyphrase Retrieval with Self-Labelling},
	journal = {Preprints.org}
}

@misc{KeyBERT,
  author       = {Maarten Grootendorst},
  title        = {KeyBERT: Minimal keyword extraction with BERT.},
  year         = 2020,
  url    = {https://github.com/MaartenGr/KeyBERT},
  note = "Software",
}

@article{downey2016blending,
  title={Blending of brain-machine interface and vision-guided autonomous robotics improves neuroprosthetic arm performance during grasping},
  author={Downey, John E and Weiss, Jeffrey M and Muelling, Katharina and Venkatraman, Arun and Valois, Jean-Sebastien and Hebert, Martial and Bagnell, J Andrew and Schwartz, Andrew B and Collinger, Jennifer L},
  journal={Journal of neuroengineering and rehabilitation},
  volume={13},
  number={1},
  pages={1--12},
  year={2016},
  publisher={BioMed Central}
}

@article{muelling2017autonomy,
  title={Autonomy infused teleoperation with application to brain computer interface controlled manipulation},
  author={Muelling, Katharina and Venkatraman, Arun and Valois, Jean-Sebastien and Downey, John E and Weiss, Jeffrey and Javdani, Shervin and Hebert, Martial and Schwartz, Andrew B and Collinger, Jennifer L and Bagnell, J Andrew},
  journal={Autonomous Robots},
  volume={41},
  pages={1401--1422},
  year={2017},
  publisher={Springer}
}

@inproceedings{lampe2014,
author = {Lampe, Thomas and Fiederer, Lukas D.J. and Voelker, Martin and Knorr, Alexander and Riedmiller, Martin and Ball, Tonio},
title = {A Brain-Computer Interface for High-Level Remote Control of an Autonomous, Reinforcement-Learning-Based Robotic System for Reaching and Grasping},
year = {2014},
isbn = {9781450321846},
publisher = {Association for Computing Machinery},
address = {New York, NY, USA},
booktitle = {Proceedings of the 19th International Conference on Intelligent User Interfaces},
pages = {83–88},
numpages = {6},
location = {Haifa, Israel},
series = {IUI '14}
}

@article{XU2019121,
title = {Shared control of a robotic arm using non-invasive brain–computer interface and computer vision guidance},
journal = {Robotics and Autonomous Systems},
volume = {115},
pages = {121-129},
year = {2019},
author = {Yang Xu and Cheng Ding and Xiaokang Shu and Kai Gui and Yulia Bezsudnova and Xinjun Sheng and Dingguo Zhang},
}

@article{cao2021brain,
  title={A brain-actuated robotic arm system using non-invasive hybrid brain--computer interface and shared control strategy},
  author={Cao, Linfeng and Li, Guangye and Xu, Yang and Zhang, Heng and Shu, Xiaokang and Zhang, Dingguo},
  journal={Journal of Neural Engineering},
  volume={18},
  number={4},
  pages={046045},
  year={2021},
  publisher={IOP Publishing}
}

@misc{gpt4,
      title={GPT-4 Technical Report}, 
      author={OpenAI},
      year={2023},
      eprint={2303.08774},
      archivePrefix={arXiv}
}

@article{cruice2003finding,
  title={Finding a focus for quality of life with aphasia: Social and emotional health, and psychological well-being},
  author={Cruice, Madeline and Worrall, Linda and Hickson, Louise and Murison, Robert},
  journal={Aphasiology},
  volume={17},
  number={4},
  pages={333--353},
  year={2003},
  publisher={Taylor \& Francis}
}

@article{xu2021human,
  title={Human Parity on CommonsenseQA: Augmenting Self-Attention with External Attention},
  author={Xu, Yichong and Zhu, Chenguang and Wang, Shuohang and Sun, Siqi and Cheng, Hao and Liu, Xiaodong and Gao, Jianfeng and He, Pengcheng and Zeng, Michael and Huang, Xuedong},
  journal={arXiv},
  year={2021}
}

@article{radford_language_nodate,
  title={Language models are unsupervised multitask learners},
  author={Radford, Alec and Wu, Jeffrey and Child, Rewon and Luan, David and Amodei, Dario and Sutskever, Ilya and others},
  journal={OpenAI blog},
  volume={1},
  number={8},
  pages={9},
  year={2019}
}

@article{krol2020cognitive,
  title={Cognitive and affective probing: a tutorial and review of active learning for neuroadaptive technology},
  author={Krol, Laurens R and Haselager, Pim and Zander, Thorsten O},
  journal={Journal of neural engineering},
  volume={17},
  number={1},
  pages={012001},
  year={2020},
  publisher={IOP Publishing}
}

@inproceedings{krol2018towards,
  title={Towards a conceptual framework for cognitive probing},
  author={Krol, Laurens R and Zander, Thorsten O},
  booktitle={Symbiotic Interaction: 6th International Workshop, Symbiotic 2017, Eindhoven, The Netherlands, December 18--19, 2017, Revised Selected Papers 6},
  pages={74--78},
  year={2018},
  organization={Springer}
}

@article{touvron2023llama,
  title={Llama: Open and efficient foundation language models},
  author={Touvron, Hugo and Lavril, Thibaut and Izacard, Gautier and Martinet, Xavier and Lachaux, Marie-Anne and Lacroix, Timoth{\'e}e and Rozi{\`e}re, Baptiste and Goyal, Naman and Hambro, Eric and Azhar, Faisal and others},
  journal={arXiv preprint arXiv:2302.13971},
  year={2023}
}

@article{wei2023jailbroken,
  title={Jailbroken: How does llm safety training fail?},
  author={Wei, Alexander and Haghtalab, Nika and Steinhardt, Jacob},
  journal={arXiv preprint arXiv:2307.02483},
  year={2023}
}

@article{perez2022ignore,
  title={Ignore previous prompt: Attack techniques for language models},
  author={Perez, F{\'a}bio and Ribeiro, Ian},
  journal={arXiv preprint arXiv:2211.09527},
  year={2022}
}

@article{greshake2023not,
  title={Not what you’ve signed up for: Compromising Real-World LLM-Integrated Applications with Indirect Prompt Injection},
  author={Greshake, Kai and Abdelnabi, Sahar and Mishra, Shailesh and Endres, Christoph and Holz, Thorsten and Fritz, Mario},
  journal={arXiv preprint arXiv:2302.12173},
  year={2023}
}

@article{lam2010relationship,
  title={The relationship of 60 disease diagnoses and 15 conditions to preference-based health-related quality of life in Ontario hospital-based long-term care residents},
  author={Lam, Jonathan MC and Wodchis, Walter P},
  journal={Medical care},
  pages={380--387},
  year={2010},
  publisher={JSTOR}
}

@article{martinez-cagigal_braincomputer_2021,
	title = {Brain–computer interfaces based on code-modulated visual evoked potentials (c-{VEP}): a literature review},
	volume = {18},
	language = {en},
	number = {6},
	journal = {Journal of Neural Engineering},
	author = {Martínez-Cagigal, Víctor and Thielen, Jordy and Santamaría-Vázquez, Eduardo and Pérez-Velasco, Sergio and Desain, Peter and Hornero, Roberto},
	month = nov,
	year = {2021},
	note = {Publisher: IOP Publishing},
	pages = {061002}
}

@Article{brainsci8070130,
AUTHOR = {Nezamfar, Hooman and Mohseni Salehi, Seyed Sadegh and Higger, Matt and Erdogmus, Deniz},
TITLE = {Code-VEP vs. Eye Tracking: A Comparison Study},
JOURNAL = {Brain Sciences},
VOLUME = {8},
YEAR = {2018},
NUMBER = {7},
ARTICLE-NUMBER = {130},
PubMedID = {29986504}
}

\pagebreak
\section*{Supplementary material}
\subsection*{EEGChat demonstration video}

A video showcasing EEGChat, the first conversational BAI, can be found at \url{https://youtu.be/uN9VEE0HJUE}\\

\subsection*{Post-experiment participant survey}
\begin{table}[htb!]
    \centering
    \begin{tabular}{p{0.35\textwidth}||p{0.1\textwidth}|p{0.1\textwidth}|p{0.1\textwidth}|p{0.1\textwidth}|p{0.1\textwidth}}
         & strongly disagree & disagree & neither agree nor disagree & agree & strongly agree \\
         \hline\hline
        I could understand the conversation partner well. &  &  &  & \textbf{1} (S2) & \textbf{4} (S1, S3, S4, \textit{S5}) \\
        \hline
        The experiment was explained well and I understood what to do during the different stages of the experiment &  &  &  &  & \textbf{5} (S1, S2, S3, S4, \textit{S5})\\
        \hline
        I was able to understand the conversation scenarios well. &  &  &  &  & \textbf{5} (S1, S2, S3, S4, \textit{S5})\\
        \hline
        The goal I needed to accomplish in each scenario was clear to me. &  &  &  &  & \textbf{5} (S1, S2, S3, S4, \textit{S5})\\
        \hline
        The functions of the special options were clear to me. &  &  &  &  & \textbf{5} (S1, S2, S3, S4, \textit{S5})\\
        \hline
        I had enough time to read all the keywords before the flickering started. &  &  &  & \textbf{2} (S4, \textit{S5}) & \textbf{3} (S1, S2, S3)\\
        \hline
        The time to read the keywords was too long.  & \textbf{1} (S1) & \textbf{1} (S4) & \textbf{3} (S2, S3, \textit{S5}) &  & \\
        \hline
        The flickering of the keywords was straining for my eyes. & \textbf{1} (S2) & \textbf{2} (S4, \textit{S5}) & \textbf{1} (S1) &  & \textbf{1} (S3) \\
        \hline
        I was always able to select the keywords I wanted to select. &  & \textbf{1} (\textit{S5}) & \textbf{1} (S2) & \textbf{2} (S1, S3) & \textbf{1} (S4)\\
        \hline
        The generated keywords were relevant to the conversation. &  &  &  & \textbf{5} (S1, S2, S3, S4, \textit{S5}) & \\
        \hline
        The keywords covered a wide range of concepts. &  & \textbf{2} (S1, S3) & \textbf{3} (S2, S4, \textit{S5}) &  &  \\
        \hline
        The keywords included options that expressed what I wanted to answer. &  &  &  & \textbf{5} (S1, S2, S3, S4, \textit{S5}) & \\
        \hline
        The generated full-sentence answers, which were read out loud by the system were appropriate for the chosen keyword.  &  &  & \textbf{1} (S2) & \textbf{4} (S1, S3, S4, \textit{S5}) & \\
        \hline
        The generated answers matched what I wanted to express  &  & \textbf{1} (\textit{S5}) & \textbf{1} (S2) & \textbf{3} (S1, S3, S4) & \\
        \hline
        The generated answers did not contain details I did not want to express  &  & \textbf{2} (S1, \textit{S5}) & \textbf{1} (S4) & \textbf{2} (S2, S3) & \\
        \hline
        The conversations I had with the experimenter made sense overall  &  &  &  & \textbf{2} (S3, \textit{S5}) & \textbf{3} (S1, S2, S4) \\
    \end{tabular}
    \caption{The results of the post-experiment participant survey. The number in each cell corresponds to the total number of votes for that option. For better readability, options without votes have been left empty. S5, due to their lower classifier performance, has been highlighted in italics for easier readability.}
    \label{tab:survey}
\end{table}

\newpage
\subsection*{Large language model prompts}
\subsubsection*{STOCK-1}
\texttt{\frenchspacing Create an answer based on the given keywords and the question:\\
Question: How are you doing?\\
Answer: I am doing great, how are you?\\
Question: Awesome, thanks. What have you been up to lately?\\
Keywords: \{keywords\}\\
Answer:}

\subsubsection*{STOCK-2}
\texttt{\frenchspacing The following is a telephone conversation. Write one sentence as the next response using the\\
keywords: \{keywords\}. Do not add any additional details.\\
Question: How are you doing?\\
Answer: I am doing great, how are you?\\
Question: Awesome, thanks. What have you been up to lately?\\
Keywords: \{keywords\}\\
Answer:}

\subsubsection*{Keyword generation prompt}
\texttt{\frenchspacing Generate N keywords that might help a speech-impaired person respond to a given question. The\\
keywords should be as short as possible and only describe one possible answer each. Provide \\
answers which are as different as possible and try to include every viewpoint in the answers.\\
For example if one of the answers is yes, also include no, and when one of the answers is good,\\
also include bad. When the question is asking for a day or time, be specific in your suggested\\
answers. In addition to suggesting answers, also provide the category of what the question is\\
asking for. For example, if the question is asking for the name of a person, the category\\
should be NAME. If the question is asking for an address or street name, the category should\\
be ADDRESS. Here are some examples:}\\
\ \\
\texttt{\frenchspacing Example 1:\\
Question: How was your day?\\
N: 6\\
Answers: 1. Good; 2. Fantastic; 3. Bad; 4. Horrible; 5. Splendid; 6. Boring\\
Category: ADJECTIVE\\
Example 2:\\
Question: How many people are living in your household?\\
N: 10\\
Answers: 1. 1; 2. 2; 3. 3; 4. 4; 5. 5; 6. 6; 7. 7; 8. 8; 9. 9; 10. 10\\
Category: NUMBER\\
Example 3:\\
Question: What is your mother's name?\\
N: 4\\
Answers: 1. Rose; 2. Mary; 3. Miriam; 4. Joanna\\
Category: NAME\\
Example 4:\\
Question: Are you hungry?\\
N: 3\\
Answers: 1. Yes; 2. No; 3. Very\\
Category: YESNO\\
\\
Question: \{question\}\\
N: \{n\_keywords\}\\
}

\subsection*{Experiment transcripts}

Full transcripts of the conversations held via the conversational BAI of all subjects. The underlined keyword was the one identified by the decoder (during training by clicking) and used to generate the full-sentence answer.

\subsubsection*{Subject 1}
\texttt{\frenchspacing
\noindent(TRAIN) Training scenario. Please follow the experimenter's instructions.\\
Q: \ Hello how are you doing today\\
KW: Good, \underline{Great}, Fine, Okay, Not Bad, Bad, Correction, More, None, Finished\\
A: \ Great, how are you?\\
Q: \ I'm good too thank you what is your favorite food\\
KW: Pizza, Sushi, Pasta, Steak, Burgers, Tacos, Correction, \underline{More}, None, Finished\\
KW: \underline{Salad}, Sandwiches, Soup, Curry, Noodles, Fish, Correction, Previous, None, Finished\\
A: \ I love salad.\\
Q: \ Now please choose correction\\
KW: Right, Correct, Accurate, Exact, Proper, True, \underline{Correction}, More, None, Finished\\
A: \ I am sorry, I misspoke earlier.\\
Q: \ Now please choose none\\
KW: None, Zero, Not any, Not one, Nothing, Not applicable, Correction, More, \underline{None}, Finished\\
A: \ I am sorry, I cannot answer this question right now.\\
Q: \ What's your favorite animal\\
KW: Dog, \underline{Cat}, Fish, Horse, Bird, Snake, Correction, More, None, Finished\\
A: \ I like cats.\\
Q: \ Please enter scenario now by selecting finished on the bottom right\\
KW: Finished, Done, Complete, Ready, Submit, Enter, Correction, More, None, \underline{Finished}\\
A: \ Thank you, good bye.\\
----------\\
\\
(UITEST) Ask what the opening hours of the museum are.\\
Q: \ Museum of modern art how can i help you\\
KW: Visit, Directions, \underline{Hours}, Admission, Exhibits, Events, Correction, More, None, Finished\\
A: \ What are the museum of modern art hours?\\
Q: \ We are open every day from 9:00 a.m. to 7:00 p.m. would you like to buy a ticket now\\
KW: Yes, No, 9:00, 10:00, 11:00, 12:00, Correction, More, None, \underline{Finished}\\
A: \ Thank you, good bye.\\
----------\\
\\
(EVAL) Make a reservation for 5 people at the pizzeria for Monday at 1pm under the name 'Anna'.\\
Q: \ Pizzeria romano how can i help you\\
KW: Order, Delivery, Takeout, \underline{Reservations}, Menu, Prices, Correction, More, None, Finished\\
A: \ I would like to make a reservation.\\
Q: \ Yes of course what day would you like to make the reservation for\\
KW: \underline{Monday}, Tuesday, Wednesday, Thursday, Friday, Saturday, Correction, More, None, Finished\\
A: \ Monday.\\
Q: \ At what time would you like to come by\\
KW: 8am, 9am, 10am, 11am, 12pm, \underline{1pm}, Correction, More, None, Finished\\
A: \ 1:00 pm.\\
Q: \ All right and how many people is the reservation for\\
KW: 1, 2, 3, 4, \underline{5}, 6, Correction, More, None, Finished\\
A: \ For 5 people.\\
Q: I am running a few minutes late; my previous meeting is running over. All right what's your name\\
KW: \underline{Anna}, Mayer, Anna Mayer, David Mayer, Laura, Oliver, Correction, More, None, Finished\\
A: \ My name is Anna.\\
Q: \ Okay anna so a reservation for monday at 1:00 p.m. for five people is that correct\\
KW: \underline{Yes}, No, Correct, Incorrect, Monday, 1:00, Correction, More, None, Finished\\
A: \ Yes.\\
Q: \ Okay anything else i can help you with\\
KW: Yes, No, Maybe, Please, Later, Soon, Correction, More, None, \underline{Finished}\\
A: \ Thank you, good bye.\\
----------\\
\\
(EVAL) Call your fitness studio and schedule a session with your trainer Sophia for Tuesday\\
morning under the name Anna Mayer.\\
Q: \ Fitness studio get fit how can i help you\\
KW: Join, Enroll, Sign up, Register, \underline{Schedule}, Prices, Correction, More, None, Finished\\
A: \ Hello, how may I help you?\\
Q: \ I'm not sure i was hoping i could help you you called the fitness studio get fit\\
KW: Yes, No, Maybe, Call, Help, Get Fit, \underline{Correction}, More, None, Finished\\
A: \ I am sorry, I misspoke earlier.\\
Q: \ Okay so how can i help you would you like to schedule a session\\
KW: \underline{Yes}, No, Maybe, Schedule, Session, Appointment, Correction, More, None, Finished\\
A: \ Yes, I would.\\
Q: \ Do you already have an account with us or would you like to sign up\\
KW: \underline{Yes}, No, Already, Sign up, Create, Existing, Correction, More, None, Finished\\
A: \ Yes, I have one.\\
Q: \ Okay what's the name of your trainer\\
KW: Anna, Mayer, Anna Mayer, David Mayer, Laura, Oliver, Correction, \underline{More}, None, Finished\\
KW: Peter, \underline{Sophia}, Tim, Marianne, Maria, Felix, Correction, Previous, None, Finished\\
A: \ Sophia.\\
Q: \ What day would you like to make an appointment for\\
KW: Monday, \underline{Tuesday}, Wednesday, Thursday, Friday, Saturday, Correction, More, None, Finished\\
A: \ Tuesday.\\
Q: \ Do you prefer morning or afternoon\\
KW: \underline{Morning}, Afternoon, Both, Neither, Early, Late, Correction, More, None, Finished\\
A: \ Morning.\\
Q: \ Is 10:00 a.m. okay\\
KW: \underline{Yes}, No, Maybe, Sure, Absolutely, Definitely, Correction, More, None, Finished\\
A: \ Yes that's fine.\\
Q: \ Okay what's your name\\
KW: Anna, Mayer, \underline{Anna Mayer}, David Mayer, Laura, Oliver, Correction, More, None, Finished\\
A: \ Anna Mayer.\\
Q: \ All right see you tuesday at 10:00 a.m. then\\
KW: Yes, No, OK, Tuesday, 10:00, 10am, Correction, More, None, \underline{Finished}\\
A: \ Thank you, good bye.\\
----------\\
\\
(EVAL) Make a doctor's appointment for next Thursday afternoon under the name 'David Mayer'.\\
Q: \ Practice doctor garcia how can i help you\\
KW: \underline{Appointment}, Prescription, Referral, Check-up, Symptom, Diagnosis, Correction, More,\\
None, Finished\\
A: \ I would like to make an appointment with Dr. Garcia.\\
Q: \ Okay what's your name\\
KW: Anna, Mayer, Anna Mayer, \underline{David Mayer}, Laura, Oliver, Correction, More, None, Finished\\
A: \ David Mayer.\\
Q: \ Okay which day would work best for you\\
KW: Monday, Tuesday, Wednesday, \underline{Thursday}, Friday, Saturday, Correction, More, None, Finished\\
A: \ I'm available on Thursday.\\
Q: \ Okay would 10:00 a.m. work\\
KW: Yes, \underline{No}, Maybe, Sure, Absolutely, Of Course, Correction, More, None, Finished\\
A: \ No, that doesn't work for me.\\
Q: \ Okay how about 3:00 p.m.\\
KW: Yes, No, Maybe, Too early, Too late, \underline{Perfect}, Correction, More, None, Finished\\
A: \ Perfect.\\
Q: \ Okay see you then\\
KW: Bye, See you, See ya, Later, Take care, Adios, Correction, More, None, \underline{Finished}\\
A: \ Thank you, good bye.\\
----------\\
\\
(EVAL) Make a reservation at the cafe for 4 people next Sunday at 4pm under the name Oliver.\\
Q: \ Cafe marie how can i help you\\
KW: Takeout, Delivery, Dine-in, Menu, Order, \underline{Reservation}, Correction, More, None, Finished\\
A: \ I have a reservation.\\
Q: \ Okay what's your name\\
KW: Anna, Mayer, Anna Mayer, David Mayer, Laura, Oliver, \underline{Correction}, More, None, Finished\\
A: \ I am sorry, I misspoke earlier.\\
Q: \ Okay how can i help you then would you like to make a\\
KW: \underline{Yes}, No, Maybe, Appointment, Purchase, Reservation, Correction, More, None, Finished\\
A: \ Yes, I would like to make a reservation.\\
Q: \ Okay for what day\\
KW: Monday, Tuesday, Wednesday, Thursday, Friday, Saturday, Correction, \underline{More}, None, Finished\\
KW: \underline{Sunday}, Today, Tomorrow, Next week, Next month, Never, Correction, Previous, None, Finished\\
A: \ Sunday, January 2nd.\\
Q: \ I'm sorry we don't take reservations that far in advance\\
KW: OK, Understand, Disappointed, Unacceptable, Acceptable, Sad, \underline{Correction}, More, None,\\
Finished\\
A: \ I am sorry, I misspoke earlier.\\
Q: \ Oh okay so which day would you like to come\\
KW: Monday, Tuesday, Wednesday, Thursday, Friday, Saturday, Correction, \underline{More}, None, Finished\\
KW: \underline{Sunday}, Morning, Afternoon, Evening, Tomorrow, Today, Correction, Previous, None, Finished\\
A: \ Sunday, January 2nd.\\
Q: \ I'm sorry like i said i can't give you a reservation for then i can offer you a reservation\\
KW: \underline{Accept}, Refuse, Later, Now, Tomorrow, Tonight, Correction, More, None, Finished\\
A: \ Okay, I accept.\\
Q: \ All right at what time would you like to come by\\
KW: 8am, 9am, 10am, 11am, 12pm, 1pm, Correction, \underline{More}, None, Finished\\
KW: 2pm, 3pm, \underline{4pm}, 5pm, 6pm, 7pm, Correction, Previous, None, Finished\\
A: \ At 4pm.\\
Q: \ Okay and for how many people\\
KW: 1, 2, 3, \underline{4}, 5, 6, Correction, More, None, Finished\\
A: \ It's going to be just me.\#\#\#\\
Q: \ Okay what's your name\\
KW: Anna, Mayer, Anna Mayer, David Mayer, Laura, Oliver, \underline{Correction}, More, None, Finished\\
A: \ I am sorry, I misspoke earlier.\\
Q: \ So how many people\\
KW: 1, 2, 3, \underline{4}, 5, 6, Correction, More, None, Finished\\
A: \ I'm sorry, I said four.\\
Q: \ Okay so four people on sunday at 4:00 p.m.\\
KW: 4, Sunday, 4:00pm, 4 people, Sunday 4pm, 4 Sunday, Correction, \underline{More}, None, Finished\\
KW: Four, Sunday 4:00, 4:00, 4pm, 4 p, Four people, Correction, Previous, \underline{None}, Finished\\
A: \ I am sorry, I cannot answer this question right now.\\
Q: \ Okay what's your name\\
KW: Anna, Mayer, \underline{Anna Mayer}, David Mayer, Laura, Oliver, Correction, More, None, Finished\\
A: \ My name is Anna Mayer.\\
Q: \ Okay i know you will be coming on sunday with four people at 4:00 p.m. is that correct\\
KW: Yes, No, Maybe, 1, 2, 3, \underline{Correction}, More, None, Finished\\
A: \ I am sorry, I misspoke earlier.\\
Q: \ Okay will you be coming at a different time\\
KW: Yes, \underline{No}, Tomorrow, Tonight, Afternoon, Evening, Correction, More, None, Finished\\
A: \ No.\#\#\#\\
Q: \ Is the number of people incorrect\\
KW: Yes, \underline{No}, Maybe, Correct, Incorrect, Could be, Correction, More, None, Finished\\
A: \ No, that number is correct.\\
Q: \ Is the day incorrect\\
KW: Yes, \underline{No}, Correct, Incorrect, Right, Wrong, Correction, More, None, Finished\\
A: \ No, the day is correct.\\
Q: \ Did i get your name wrong\\
KW: \underline{Yes}, No, Maybe, Correct, Incorrect, Yes, it's, Correction, More, None, Finished\\
A: \ Yes.\\
Q: \ What is your name\\
KW: Anna, Mayer, Anna Mayer, David Mayer, Laura, \underline{Oliver}, Correction, More, None, Finished\\
A: \ My name is Oliver.\\
Q: \ Okay oliver then i will see you on sunday at 4:00\\
KW: Yes, No, OK, Sunday, 4, 4:00, Correction, More, None, \underline{Finished}\\
A: \ Thank you, good bye.\\
----------\\
\\
(EVAL) Order a taxi to 15 Flowerstreet for Wednesday at 3pm for 2 people.\\
Q: \ Taxi service jones how can i help you\\
KW: \underline{Book}, Reserve, Call, Request, Cancel, Change, Correction, More, None, Finished\\
A: \ Can you book a taxi to the airport?\\
Q: \ Yes of course for which day\\
KW: Monday, Tuesday, Wednesday, Thursday, Friday, Saturday, \underline{Correction}, More, None, Finished\\
A: \ I am sorry, I misspoke earlier.\\
Q: \ Okay how can i help you as a taxi service\\
KW: Pickup, Dropoff, Airport, Home, Hotel, Destination, Correction, \underline{More}, None, Finished\\
KW: Ride, Fare, Time, Now, Soon, Later, Correction, \underline{Previous}, None, Finished\\
KW: Pickup, Dropoff, Airport, Home, Hotel, \underline{Destination}, Correction, More, None, Finished\\
A: \ I need a taxi to take me to the airport.\\
Q: \ Yes okay where would you like to be picked up\\
KW: Home, Work, School, Airport, Train Station, Bus Stop, \underline{Correction}, More, None, Finished\\
A: \ I am sorry, I misspoke earlier.\\
Q: \ Do you want to go to the airport or somewhere else\\
KW: Airport, \underline{Elsewhere}, Yes, No, Maybe, Definitely, Correction, More, None, Finished\\
A: \ Elsewhere.\\
Q: \ Okay no problem where would you like to be picked up\\
KW: \underline{Home}, Office, School, Park, Mall, Restaurant, Correction, More, None, Finished\\
A: \ I am at home.\\
Q: \ And what's your address\\
KW: 15 Flowerstreet, 7 Southroad, 104 Mainstreet, \underline{28 Bumblebee Lane}, 56 Park Avenue,\\
7 St. Michael Road, Correction, More, None, Finished\\
A: \ 28 Bumblebee Lane.\\
Q: \ For which day would you like to schedule your pickup\\
KW: Monday, Tuesday, \underline{Wednesday}, Thursday, Friday, Saturday, Correction, More, None, Finished\\
A: \ Wednesday.\\
Q: \ And at what time\\
KW: 12am, 12pm, 1am, 1pm, 2am, 2pm, Correction, \underline{More}, None, Finished\\
KW: 3am, \underline{3pm}, 4am, 4pm, 5am, 5pm, Correction, Previous, None, Finished\\
A: \ Can you pick me up at 3 o'clock?\\
Q: \ Yes of course will you require an extra big taxi or is a regular car okay\\
KW: Big, \underline{Regular}, Extra Big, Small, Medium, Yes, Correction, More, None, Finished\\
A: \ A regular taxi will be okay.\\
Q: \ Okay i will drive over to come pick you up then\\
KW: Yes, No, Sure, Alright, OK, Sounds Good, \underline{Correction}, More, None, Finished\\
A: \ I am sorry, I misspoke earlier.\\
Q: \ Do you do you require a big taxi\\
KW: Yes, \underline{No}, Small, Medium, Large, Extra Large, Correction, More, None, Finished\\
A: \ No, a regular taxi will be fine.\\
Q: \ And the pickup time\\
KW: 8am, 8:15am, 8:30am, 8:45am, 9am, 9:15am, Correction, \underline{More}, None, Finished\\
KW: 9:30am, 9:45am, 10am, 10:15am, 10:30am, 10:45am, Correction, Previous, \underline{None}, Finished\\
A: \ I am sorry, I cannot answer this question right now.\\
Q: \ Okay what day am i picking you up\\
KW: Monday, Tuesday, \underline{Wednesday}, Thursday, Friday, Saturday, Correction, More, None, Finished\\
A: \ Wednesday.\\
Q: \ All right and at what time should i come\\
KW: 8am, 9am, 10am, 11am, 12pm, 1pm, Correction, \underline{More}, None, Finished\\
KW: 2pm, \underline{3pm}, 4pm, 5pm, 6pm, 7pm, Correction, Previous, None, Finished\\
A: \ You can pick me up at 3 o'clock.\\
Q: \ So wednesday at 3:00 at 28 bumblebee lane is that correct\\
KW: Yes, \underline{No}, Wednesday, 3:00, 28, Bumblebee, Correction, More, None, Finished\\
A: \ No, that's the wrong address.\\
Q: \ So what is your address\\
KW: \underline{15 Flowerstreet}, 7 Southroad, 104 Mainstreet, 28 Bumblebee Lane, 56 Park Avenue,\\
7 St. Michael Road, Correction, More, None, Finished\\
A: \ 15 Flowerstreet.\\
Q: \ Okay i would rather will be there\\
KW: Yes, No, Maybe, Certainly, Absolutely, Of course, Correction, More, None, \underline{Finished}\\
A: \ Thank you, good bye.\\
----------\\
\\
(EVAL) Make an appointment at the hair salon to have your hair cut on Tuesday afternoon under\\
the name Anna.\\
Q: \ Leslie's hair salon how can i help you\\
KW: Book, \underline{Appointment}, Directions, Hours, Price, Promotion, Correction, More, None, Finished\\
A: \ Hi, I'm calling about my appointment today.\\
Q: \ Okay what's your name\\
KW: Anna, Mayer, Anna Mayer, David Mayer, Laura, Oliver, \underline{Correction}, More, None, Finished\\
A: \ I am sorry, I misspoke earlier.\\
Q: \ Oh okay would you like to make an make an appointment\\
KW: \underline{Yes}, No, Sure, Later, Now, Tomorrow, Correction, More, None, Finished\\
A: \ Yes, my name is Sara.\\
Q: \ Okay what day would you like to come by\\
KW: Monday, Tuesday, Wednesday, Thursday, Friday, Saturday, \underline{Correction}, More, None, Finished\\
A: \ I am sorry, I misspoke earlier.\\
Q: \ Okay what's your name\\
KW: \underline{Anna}, Mayer, Anna Mayer, David Mayer, Laura, Oliver, Correction, More, None, Finished\\
A: \ Hi, I'm calling about my appointment today. My name is Anna.\\
Q: \ I don't have an appointment for an anna today would you like to make one\\
KW: \underline{Yes}, No, Maybe, Later, Soon, Today, Correction, More, None, Finished\\
A: \ Yes, I would.\\
Q: \ Okay which day would you like to come\\
KW: Monday, \underline{Tuesday}, Wednesday, Thursday, Friday, Saturday, Correction, More, None, Finished\\
A: \ Tuesday is fine.\\
Q: \ Okay have open slots at 10 a.m. or 4:00 p.m. would one of those work for you\\
KW: 10am, \underline{4pm}, Yes, No, Maybe, Absolutely, Correction, More, None, Finished\\
A: \ 4:00 p.m. That would work for me.\\
Q: \ Is the apartment for cutting for coloring or styling\\
KW: \underline{Cutting}, Coloring, Styling, Both, Neither, Both cutting and coloring, Correction, More,\\
None, Finished\\
A: \ I need a haircut.\\
Q: \ Okay is there anything else i can help you with\\
KW: Yes, \underline{No}, Maybe, More Info, Explain, Clarify, Correction, More, None, Finished\\
A: \ No, that's all.\\
Q: \ Okay and i'll see you then\\
KW: Yes, \underline{No}, Sure, Alright, Fine, Definitely, Correction, More, None, Finished\\
A: \ No, that's all.\\
Q: \ Okay bye\\
KW: Bye, See ya, Adios, Hasta luego, Ciao, Ta ta, Correction, More, None, \underline{Finished}\\
A: \ Thank you, good bye.\\
}

\subsubsection*{Subject 2}
\texttt{\frenchspacing
\noindent(TRAIN) Training scenario. Please follow the experimenter's instructions.\\
Q: \ Hello how are you doing today\\
KW: \underline{Great}, Good, Bad, Terrible, Well, Poorly, Correction, More, None, Finished\\
A: \ I'm great.\\
Q: \ What is your favorite food\\
KW: Pizza, \underline{Pasta}, Burgers, Salad, Steak, Sushi, Correction, More, None, Finished\\
A: \ Pasta.\\
Q: \ Now please choose correction\\
KW: Right, Correct, Accurate, Precise, Exact, Faultless, \underline{Correction}, More, None, Finished\\
A: \ I am sorry, I misspoke earlier.\\
Q: \ Oh okay what is your favorite food\\
KW: Pizza, Pasta, Sushi, Steak, \underline{Tacos}, Burgers, Correction, More, None, Finished\\
A: \ Tacos.\\
Q: \ From the options on the right side please first select more and then select\\
KW: Yes, No, More, Less, Top, Bottom, Correction, \underline{More}, None, Finished\\
KW: Left, Right, Up, Down, Front, Back, Correction, Previous, \underline{None}, Finished\\
A: \ I am sorry, I cannot answer this question right now.\\
Q: \ What is your name\\
KW: Anna, Mayer, Anna Mayer, David Mayer, Laura, Oliver, Correction, \underline{More}, None, Finished\\
KW: Peter, Sophia, Tim, Marianne, Maria, Felix, Correction, Previous, \underline{None}, Finished\\
A: \ I am sorry, I cannot answer this question right now.\\
Q: \ Please end the scenario now by selecting finished on the bottom right\\
KW: Finished, End, Stop, Done, Exit, Complete, Correction, More, None, \underline{Finished}\\
A: \ Thank you, good bye.\\
----------\\
\\
(UITEST) Ask what the opening hours of the museum are.\\
Q: \ Museum of modern art how can i help you\\
KW: Visit, \underline{Hours}, Location, Directions, Exhibits, Admission, Correction, More, None, Finished\\
A: \ I was wondering what your hours were.\\
Q: \ We are open every day from 9:00 a.m. to 7:00 p.m.\\
KW: 9am, 10am, 11am, 12pm, 1pm, 2pm, Correction, More, None, \underline{Finished}\\
A: \ Thank you, good bye.\\
----------\\
\\
(EVAL) Make a reservation for 2 people at the pizzeria for Saturday at 8pm under the name 'Anna'.\\
KW: Order, Delivery, Takeout, Pickup, Menu, Location, Correction, \underline{More}, None, Finished\\
KW: Hours, \underline{Reservations}, Catering, Specials, Gift Cards, Phone Number, Correction, Previous,\\
None, Finished\\
A: \ I would like to make a reservation.\\
Q: \ Okay what day is the reservation for\\
KW: Monday, Tuesday, Wednesday, Thursday, Friday, \underline{Saturday}, Correction, More, None, Finished\\
A: \ It's for Saturday.\\
Q: \ Okay for how many people\\
KW: 1, \underline{2}, 3, 4, 5, 6, Correction, More, None, Finished\\
A: \ It's for 2 people.\\
Q: \ What time are you coming\\
KW: 8am, 9am, 10am, 11am, 12pm, 1pm, Correction, \underline{More}, None, Finished\\
KW: 2pm, 3pm, 4pm, 5pm, 6pm, 7pm, Correction, Previous, \underline{None}, Finished\\
A: \ I am sorry, I cannot answer this question right now.\\
Q: \ Is the table for lunch or for dinner\\
KW: Lunch, \underline{Dinner}, Both, Neither, Morning, Afternoon, Correction, More, None, Finished\\
A: \ I would like to make it for dinner.\\
Q: \ Okay we have an open table at 7:00 at 8:00 or at 9:00 would one of those work\\
KW: 7:00, 7:30, \underline{8:00}, 8:30, 9:00, 9:30, Correction, More, None, Finished\\
A: \ 8:00 is perfect.\\
Q: \ Okay what's your name\\
KW: \underline{Anna}, Mayer, Anna Mayer, David Mayer, Laura, Oliver, Correction, More, None, Finished\\
A: \ This is Anna.\\
Q: \ Okay and i'll see you then\\
KW: Yes, No, OK, Sure, Certainly, Absolutely, Correction, More, None, \underline{Finished}\\
A: \ Thank you, good bye.\\
----------\\
\\
(EVAL) Call your fitness studio and schedule a session with your trainer Sophia for Friday\\
morning under the name David Mayer.\\
KW: Join, Schedule, Class, Cost, Membership, Location, Correction, \underline{More}, None, Finished\\
KW: Trainers, Hours, Facilities, Sign-up, Equipment, Programs, Correction, \underline{Previous}, None,\\
Finished\\
KW: Join, Schedule, \underline{Class}, Cost, Membership, Location, Correction, More, None, Finished\\
A: \ I'm interested in taking a class. How can I help you?\\
Q: \ First of all i need to know if you're already a member or would you like to sign up\\
KW: \underline{Member}, Sign up, Yes, No, Already, Join, Correction, More, None, Finished\\
A: \ I'm a member.\\
Q: \ Okay what's your name\\
KW: Anna, Mayer, Anna Mayer, \underline{David Mayer}, Laura, Oliver, Correction, More, None, Finished\\
A: \ David Mayer.\\
Q: \ All right and what's your trainer's name\\
KW: Anna, Mayer, Anna Mayer, David Mayer, Laura, Oliver, Correction, \underline{More}, None, Finished\\
KW: Peter, \underline{Sophia}, Tim, Marianne, Maria, Felix, Correction, Previous, None, Finished\\
A: \ Sophia.\\
Q: \ And what day would you like to make an appointment for\\
KW: Monday, Tuesday, Wednesday, Thursday, \underline{Friday}, Saturday, Correction, More, None, Finished\\
A: \ Friday.\\
Q: \ Can you prefer morning or afternoon\\
KW: \underline{Morning}, Afternoon, Both, Neither, Early, Late, Correction, More, None, Finished\\
A: \ Morning.\\
Q: \ Is 10:00 a.m. okay\\
KW: \underline{Yes}, No, Sure, Absolutely, Definitely, Not really, Correction, More, None, Finished\\
A: \ Yes, that's fine.\\
Q: \ Okay anything else i can do for you mr mayor\\
KW: \underline{No}, Yes, Maybe, Help, Assist, Support, Correction, More, None, Finished\\
A: \ No, that's all.\\
Q: \ Okay see you soon\\
KW: Bye, Goodbye, See you, See ya, See you later, Later, Correction, More, None, \underline{Finished}\\
A: \ Thank you, good bye.\\
----------\\
\\
(EVAL) Make a doctor's appointment for next Wednesday afternoon under the name 'David Mayer'.\\
Q: \ Practice dr garcia how can i help you\\
KW: Schedule, \underline{Appointment}, Consultation, Payment, Insurance, Tests, Correction, More, None,\\
Finished\\
A: \ I have an appointment. Can I have the address?\\
Q: \ Yes our address is school street number one\\
KW: Yes, No, Correct, Incorrect, School, Street, \underline{Correction}, More, None, Finished\\
A: \ I am sorry, I misspoke earlier.\\
Q: \ Oh okay so this is practice of dr garcia how can i help you\\
KW: Question, Request, Problem, Advice, Answer, Information, Correction, \underline{More}, None, Finished\\
KW: Suggestion, Assistance, Help, Inquiry, Clarification, Guidance, Correction, \underline{Previous}, None,\\
Finished\\
KW: Question, \underline{Request}, Problem, Advice, Answer, Information, Correction, More, None, Finished\\
A: \ I'd like to request an appointment.\\
Q: \ What's your name\\
KW: Anna, Mayer, Anna Mayer, \underline{David Mayer}, Laura, Oliver, Correction, More, None, Finished\\
A: \ That's me, David Mayer.\\
Q: \ Which day would work best for you\\
KW: Monday, Tuesday, \underline{Wednesday}, Thursday, Friday, Saturday, Correction, More, None, Finished\\
A: \ I was hoping for Wednesday.\\
Q: \ Okay it's 10:00 a.m. okay\\
KW: 10:00, 10am, Ten, 10, Ten a, Ten in the morning, Correction, \underline{More}, None, Finished\\
KW: Ten o'clock, Ten o'clock in the morning, Ten o'clock a, Ten o'clock am, Yes, \underline{No}, Correction, Previous, None, Finished\\
A: \ No, that's not okay.\\
Q: \ How about 3:00 p.m.\\
KW: \underline{Yes}, No, Good, Bad, Fine, Not good, Correction, More, None, Finished\\
A: \ Yes, that would be perfect.\\
Q: \ Okay see you then\\
KW: \underline{Bye}, See you, Adios, Goodbye, Ciao, Hasta luego, Correction, More, None, Finished\\
A: \ Bye.\\
Q: \ All right see you\\
KW: OK, Bye, See ya, Ciao, Adios, Take care, Correction, More, \underline{None}, Finished\\
A: \ I am sorry, I cannot answer this question right now.\\
Q: \ Can i help you with anything else\\
KW: Yes, No, Maybe, Certainly, Absolutely, Certainly Not, Correction, More, None, \underline{Finished}\\
A: \ Thank you, good bye.\\
----------\\
\\
(EVAL) Make a reservation at the cafe for 5 people next Wednesday at 3pm under the name Laura.\\
KW: Order, Delivery, Menu, Pickup, Specials, Drinks, Correction, \underline{More}, None, Finished\\
KW: Hours, Location, Payment, \underline{Reservations}, Catering, Takeaway, Correction, Previous, None,\\
Finished\\
A: \ I would like to make a reservation.\\
Q: \ What day would you like to come\\
KW: Monday, Tuesday, \underline{Wednesday}, Thursday, Friday, Saturday, Correction, More, None, Finished\\
A: \ I would like to come on Wednesday.\\
Q: \ At what time would you come by\\
KW: 8am, 9am, 10am, 11am, 12pm, 1pm, Correction, \underline{More}, None, Finished\\
KW: 2pm, \underline{3pm}, 4pm, 5pm, 6pm, 7pm, Correction, Previous, None, Finished\\
A: \ Would 3pm be okay?\\
Q: \ Yes of course how many people will you be\\
KW: 1, 2, 3, 4, \underline{5}, 6, Correction, More, None, Finished\\
A: \ We will be 5.\\
Q: \ All right thank you what's your name\\
KW: Anna, Mayer, Anna Mayer, David Mayer, \underline{Laura}, Oliver, Correction, More, None, Finished\\
A: \ This is Laura.\\
Q: \ Okay laura see you then\\
KW: Yes, No, Alright, Sure, Sounds good, Got it, Correction, More, None, \underline{Finished}\\
A: \ Thank you, good bye.\\
----------\\
\\
(EVAL) Order a taxi to 104 Mainstreet for Friday at 11am for 2 people.\\
Q: \ Taxi service jones how can i help you\\
KW: Schedule, \underline{Book}, Cancel, Price, Location, Hours, Correction, More, None, Finished\\
A: \ I would like to book a taxi for tomorrow.\\
Q: \ Okay what's your address\\
KW: 15 Flowerstreet, 7 Southroad, 104 Mainstreet, 28 Bumblebee Lane, 56 Park Avenue,\\
7 St. Michael Road, \underline{Correction}, More, None, Finished\\
A: \ I am sorry, I misspoke earlier.\\
Q: \ Is the taxi not for tomorrow\\
KW: Yes, \underline{No}, Maybe, Not sure, Definitely, Certainly, Correction, More, None, Finished\\
A: \ No, it's for tomorrow.\\
Q: \ Which day is the taxi for\\
KW: Monday, Tuesday, Wednesday, Thursday, \underline{Friday}, Saturday, Correction, More, None, Finished\\
A: \ Friday.\\
Q: \ All right friday then and how many people\\
KW: 1, \underline{2}, 3, 4, 5, 6, Correction, More, None, Finished\\
A: \ I will be two people.\\
Q: \ And what's your address\\
KW: 15 Flowerstreet, 7 Southroad, \underline{104 Mainstreet}, 28 Bumblebee Lane, 56 Park Avenue,\\
7 St. Michael Road, Correction, More, None, Finished\\
A: \ 104 Mainstreet.\\
Q: \ At what time should the driver come by\\
KW: 9am, 10am, \underline{11am}, 12pm, 1pm, 2pm, Correction, More, None, Finished\\
A: \ Around 11am.\\
Q: \ Okay friday at 11:00 a.m. is that correct\\
KW: \underline{Yes}, No, Friday, 11, 11am, 11:00, Correction, More, None, Finished\\
A: \ Yes that's correct.\\
Q: \ Okay thank you\\
KW: You're welcome, No problem, My pleasure, Sure, Anytime, No worries, Correction, More, None,\\
\underline{Finished}\\
A: \ Thank you, good bye.\\
----------\\
\\
(EVAL) Make an appointment at the hair salon to have your hair cut on Thursday morning under\\
the name Tim.\\
Q: \ Leslie's hair salon how can i help you\\
KW: Book, \underline{Appointment}, Hours, Prices, Services, Directions, Correction, More, None, Finished\\
A: \ Do you have an appointment?\\
Q: \ Yes we have open appointments which day would you like to come by\\
KW: Monday, Tuesday, Wednesday, \underline{Thursday}, Friday, Saturday, Correction, More, None, Finished\\
A: \ Thursday would be good.\\
Q: \ Open appointments at 10:00 a.m. or at 3:00 p.m. would one of those work for you\\
KW: \underline{10AM}, 3PM, Yes, No, Maybe, Definitely, Correction, More, None, Finished\\
A: \ 10AM would be fine.\\
Q: \ Is the appointment for cutting coloring or styling\\
KW: \underline{Cutting}, Coloring, Styling, Both, Neither, Cut, Correction, More, None, Finished\\
A: \ Cutting \\
Q: \ All right what's your name\\
KW: Anna, Mayer, Anna Mayer, David Mayer, Laura, Oliver, Correction, \underline{More}, None, Finished\\
KW: Peter, Sophia, \underline{Tim}, Marianne, Maria, Felix, Correction, Previous, None, Finished\\
A: \ Tim.\\
Q: \ Okay see you then tim\\
KW: Bye, See ya, Ciao, Later, Adios, Hasta luego, Correction, More, None, \underline{Finished}\\
A: \ Thank you, good bye.\\
}

\subsubsection*{Subject 3}
\texttt{\frenchspacing
\noindent(TRAIN) Training scenario. Please follow the experimenter's instructions.\\
Q: \ Hello how are you doing today\\
KW: \underline{Good}, Great, Fine, Not Bad, Okay, All Right, Correction, More, None, Finished\\
A: \ I'm good.\\
Q: \ What is your favorite food\\
KW: Pizza, Steak, Salad, Pasta, Sushi, \underline{Ramen}, Correction, More, None, Finished\\
A: \ Ramen.\\
Q: \ Now please select correction again with looking\\
KW: Right, Correct, Accurate, Precise, Exact, Proper, \underline{Correction}, More, None, Finished\\
A: \ I am sorry, I misspoke earlier.\\
Q: \ Oh okay what's your favorite food\\
KW: Pizza, Pasta, \underline{Burger}, Steak, Sushi, Tacos, Correction, More, None, Finished\\
A: \ I love burgers.\\
Q: \ From the options on the right please select more and afterwards select none\\
KW: More, Less, All, None, Some, Few, Correction, \underline{More}, None, Finished\\
KW: Majority, Minority, Entire, Partial, Full, Empty, Correction, Previous, \underline{None}, Finished\\
A: \ I am sorry, I cannot answer this question right now.\\
Q: \ Now please end the scenario by selecting finished on the bottom right\\
KW: Done, Finished, Complete, Over, Ended, Wrapped, Correction, More, None, \underline{Finished}\\
A: \ Thank you, good bye.\\
----------\\
\\
(UITEST) Ask what the opening hours of the museum are.\\
KW: Understand, Learn, Discover, Explore, Analyze, Research, Correction, \underline{More}, None, Finished\\
KW: Summarize, Compare, Critique, Examine, Assess, Evaluate, Correction, Previous, \underline{None},\\
Finished\\
A: \ I am sorry, I cannot answer this question right now.\\
Q: \ You've called the museum of modern art what can i help you with\\
KW: Visit, Tickets, Information, Directions, \underline{Hours}, Exhibitions, Correction, More, None,\\
Finished\\
A: \ Do you have the hours?\\
Q: \ Usher we're open every day from 9:00 a.m. to 7:00 p.m.\\
KW: 9am, 10am, 11am, 12pm, 1pm, 2pm, Correction, More, None, \underline{Finished}\\
A: \ Thank you, good bye.\\
----------\\
\\
(EVAL) Make a reservation for 6 people at the pizzeria for Monday at 2pm under the name \\
'David Mayer'.\\
Q: \ Pizzeria romano how can i help you\\
KW: Delivery, Takeout, Menu, Hours, Location, \underline{Reservations}, Correction, More, None, Finished\\
A: \ Hi, may I have a reservation?\\
Q: \ Yes of course what day is the reservation for\\
KW: \underline{Monday}, Tuesday, Wednesday, Thursday, Friday, Saturday, Correction, More, None, Finished\\
A: \ Monday.\\
Q: \ At what time would you like to come by\\
KW: 8am, 9am, 10am, 11am, 12pm, 1pm, Correction, \underline{More}, None, Finished\\
KW: \underline{2pm}, 3pm, 4pm, 5pm, 6pm, 7pm, Correction, Previous, None, Finished\\
A: \ 2pm.\\
Q: \ Okay and how many people\\
KW: 1, 2, 3, 4, 5, \underline{6}, Correction, More, None, Finished\\
A: \ There will be 6 of us.\\
Q: \ All right what's your name\\
KW: Anna, Mayer, Anna Mayer, \underline{David Mayer}, Laura, Oliver, Correction, More, None, Finished\\
A: \ My name is David Mayer.\\
Q: \ Okay anything else i can help you with\\
KW: Yes, No, Maybe, Information, Advice, Assistance, Correction, More, None, \underline{Finished}\\
A: \ Thank you, good bye.\\
----------\\
\\
(EVAL) Call your fitness studio and schedule a session with your trainer Sophia for Friday\\
afternoon under the name Anna Mayer.\\
Q: \ Fitness studio get fit how can i help you\\
KW: Join, Classes, Prices, \underline{Schedule}, Location, Facilities, Correction, More, None, Finished\\
A: \ I'd like to schedule a fitness assessment.\\
Q: \ Okay are you already a member with us\\
KW: \underline{Yes}, No, Maybe, Possibly, Certainly, Absolutely, Correction, More, None, Finished\\
A: \ Yes, I am.\\
Q: \ All right what's the name of your trainer\\
KW: Anna, Mayer, Anna Mayer, David Mayer, Laura, Oliver, Correction, \underline{More}, None, Finished\\
KW: Peter, \underline{Sophia}, Tim, Marianne, Maria, Felix, Correction, Previous, None, Finished\\
A: \ Sophia.\\
Q: \ Okay and what's your name\\
KW: Anna, Mayer, \underline{Anna Mayer}, David Mayer, Laura, Oliver, Correction, More, None, Finished\\
A: \ Anna Mayer.\\
Q: \ All right what day would you like to make an appointment\\
KW: Monday, Tuesday, Wednesday, Thursday, \underline{Friday}, Saturday, Correction, More, None, Finished\\
A: \ I'd like to do it on Friday.\\
Q: \ Do you prefer morning or afternoon\\
KW: Morning, \underline{Afternoon}, Both, Neither, Early, Late, Correction, More, None, Finished\\
A: \ Afternoon.\\
Q: \ Is 3:00 p.m. okay\\
KW: \underline{Yes}, No, Sure, Alright, Absolutely, Definitely, Correction, More, None, Finished\\
A: \ Yes.\\
Q: \ Okay i haven't booked is there anything else i can help you with\\
KW: Yes, \underline{No}, Maybe, Later, Not now, Not yet, Correction, More, None, Finished\\
A: \ No, that's all.\\
Q: \ Okay see you then\\
KW: Bye, See you, See ya, See you later, Bye bye, Later, Correction, More, None, \underline{Finished}\\
A: \ Thank you, good bye.\\
----------\\
\\
(EVAL) Make a doctor's appointment for next Friday afternoon under the name 'David Mayer'.\\
Q: \ Practice doctor garcia how can i help you\\
KW: Question, \underline{Appointment}, Prescription, Advice, Treatment, Referral, Correction, More, None,\\
Finished\\
A: \ I want to make an appointment.\\
Q: \ What's your name\\
KW: Anna, Mayer, Anna Mayer, \underline{David Mayer}, Laura, Oliver, Correction, More, None, Finished\\
A: \ David Mayer.\\
Q: \ Which day would work best for you\\
KW: Monday, Tuesday, Wednesday, Thursday, \underline{Friday}, Saturday, Correction, More, None, Finished\\
A: \ Friday?\\
Q: \ What is 10 okay\\
KW: Acceptable, Sufficient, Adequate, Tolerable, All right, Passable, Correction, More, \underline{None},\\
Finished\\
A: \ I am sorry, I cannot answer this question right now.\\
Q: \ Is an appointment at 10 a.m. all right with you\\
KW: Yes, \underline{No}, Absolutely, Of course, Certainly, Definitely, Correction, More, None, Finished\\
A: \ Not really.\\
Q: \ Okay how about 3:00 p.m.\\
KW: \underline{Yes}, No, Maybe, 0:00, 1:00, 2:00, Correction, More, None, Finished\\
A: \ Yes that's fine.\\
Q: \ All right see you then\\
KW: Bye, Cya, Take care, See ya, See you later, Goodbye, Correction, More, None, \underline{Finished}\\
A: \ Thank you, good bye.\\
----------\\
\\
(EVAL) Make a reservation at the cafe for 5 people next Monday at 3pm under the name Anna.\\
Q: \ Cafe marie how can i help you\\
KW: Order, Menu, Delivery, Takeout, \underline{Reservations}, Hours, Correction, More, None, Finished\\
A: \ I would like to make a reservation.\\
Q: \ Okay at what time are you coming\\
KW: 9am, 10am, 11am, 12pm, 1pm, 2pm, Correction, \underline{More}, None, Finished\\
KW: \underline{3pm}, 4pm, 5pm, 6pm, 7pm, 8pm, Correction, Previous, None, Finished\\
A: \ I will be there at 3pm\\
Q: \ And on which day\\
KW: \underline{Monday}, Tuesday, Wednesday, Thursday, Friday, Saturday, Correction, More, None, Finished\\
A: \ Monday.\\
Q: \ How many people is the reservation for\\
KW: 1, 2, 3, 4, \underline{5}, 6, Correction, More, None, Finished\\
A: \ 5 people.\\
Q: \ All right what's your name\\
KW: \underline{Anna}, Mayer, Anna Mayer, David Mayer, Laura, Oliver, Correction, More, None, Finished\\
A: \ My name is Anna.\\
Q: \ All right i've knocked it down thank you\\
KW: No problem, You're welcome, Sure, Glad to help, Anytime, No worries, Correction, More, None, \underline{Finished}\\
A: \ Thank you, good bye.\\
----------\\
\\
(EVAL) Order a taxi to 15 Flowerstreet for Thursday at 5pm for 5 people.\\
Q: \ Taxi service jones how can i help you\\
KW: Book, Schedule, Cancel, Driver, \underline{Ride}, Pickup, Correction, More, None, Finished\\
A: \ I would like to schedule a ride.\\
Q: \ Okay for which day\\
KW: Monday, Tuesday, Wednesday, \underline{Thursday}, Friday, Saturday, Correction, More, None, Finished\\
A: \ Thursday.\\
Q: \ At what time would you like to be picked up\\
KW: 9am, 9:30am, 10am, 10:30am, 11am, 11:30am, Correction, \underline{More}, None, Finished\\
KW: 12pm, 12:30pm, 1pm, 1:30pm, 2pm, 2:30pm, Correction, Previous, \underline{None}, Finished\\
A: \ I am sorry, I cannot answer this question right now.\\
Q: \ What's your address\\
KW: \underline{15 Flowerstreet}, 7 Southroad, 104 Mainstreet, 28 Bumblebee Lane, 56 Park Avenue,\\
7 St. Michael Road, Correction, More, None, Finished\\
A: \ 15 Flowerstreet\\
Q: \ And when should our driver be there\\
KW: 8am, 9am, 10am, 11am, 12pm, 1pm, Correction, \underline{More}, None, Finished\\
KW: 2pm, 3pm, 4pm, \underline{5pm}, 6pm, 7pm , Correction, Previous, None, Finished\\
A: \ 5pm\\
Q: \ Do you require an extra big taxi or is a regular car okay\\
KW: \underline{Big}, Regular, Yes, No, Extra, Small, Correction, More, None, Finished\\
A: \ A big taxi is fine\\
Q: \ Okay your reservation is confirmed\\
KW: Great, Perfect, Wonderful, Fantastic, Superb, Excellent, Correction, More, None, \underline{Finished}\\
A: \ Thank you, good bye.\\
----------\\
\\
(EVAL) Make an appointment at the hair salon to have your hair styled on Wednesday afternoon\\
under the name Anna.\\
Q: \ Leslie's hair salon how can i help you\\
KW: Book, \underline{Appointment}, Cut, Color, Style, Products, Correction, More, None, Finished\\
A: \ I want to make an appointment.\\
Q: \ What day would you like to come by\\
KW: Monday, Tuesday, \underline{Wednesday}, Thursday, Friday, Saturday, Correction, More, None, Finished\\
A: \ Wednesday.\\
Q: \ Okay we have open slots at 9:00 a.m. or 4:00 p.m. would one of those work for you\\
KW: 9am, \underline{4pm}, Yes, No, Sure, Not Sure, Correction, More, None, Finished\\
A: \ The 4:00 p.m. slot works best for me.\\
Q: \ Okay is the apartment for cutting coloring or styling\\
KW: Cutting, Coloring, \underline{Styling}, Both, Neither, All, Correction, More, None, Finished\\
A: \ Styling all of the above.\\
Q: \ So styling\\
KW: Fabulous, Trendy, Chic, Modern, Classic, Elegant, Correction, \underline{More}, None, Finished\\
KW: Funky, Stylish, Daring, Outdated, Awful, Boring, \underline{Correction}, Previous, None, Finished\\
A: \ I am sorry, I misspoke earlier.\\
Q: \ Okay is it for cutting for coloring or for styling\\
KW: Cutting, Coloring, \underline{Styling}, Both, Neither, All, Correction, More, None, Finished\\
A: \ I was calling about styling.\\
Q: \ Okay perfect and what's your name\\
KW: \underline{Anna}, Mayer, Anna Mayer, David Mayer, Laura, Oliver, Correction, More, None, Finished\\
A: \ Anna.\\
Q: \ Okay i know see you then\\
KW: Yes, No, Bye, See you, Sure, Alright, Correction, More, None, \underline{Finished}\\
A: \ Thank you, good bye.\\
}

\subsubsection*{Subject 4}
\texttt{\frenchspacing
\noindent(TRAIN) Training scenario. Please follow the experimenter's instructions.\\
Q: \ Hello how are you doing today\\
KW: Good, \underline{Great}, OK, Bad, Terrible, Fantastic, Correction, More, None, Finished\\
A: \ Great, how are you?\\
Q: \ I'm also good thanks what's your favorite food\\
KW: Pizza, Burgers, Tacos, Sushi, \underline{Pasta}, Steak, Correction, More, None, Finished\\
A: \ Pasta.\\
Q: \ Now please choose correction on the right side\\
KW: Right, Left, Above, Below, Up, Down, \underline{Correction}, More, None, Finished\\
A: \ I am sorry, I misspoke earlier.\\
Q: \ Oh okay what's your favorite food\\
KW: \underline{Pizza}, Pasta, Steak, Salad, Sushi, Tacos, Correction, More, None, Finished\\
A: \ Pizza.\\
Q: \ From the options on the right side please select not please let more and then none\\
KW: None, One, Two, Three, Four, Five, Correction, \underline{More}, None, Finished\\
KW: Six, Seven, Eight, Nine, Ten, More, Correction, Previous, \underline{None}, Finished\\
A: \ I am sorry, I cannot answer this question right now.\\
Q: \ Please enter scenario now by selecting finished on the right side\\
KW: Started, Initiated, Begun, Commenced, Finished, Ended, Correction, More, None, \underline{Finished}\\
A: \ Thank you, good bye.\\
----------\\
\\
(UITEST) Ask what the opening hours of the museum are.\\
Q: \ Museum of modern art how can i help you\\
KW: Visit, \underline{Hours}, Directions, Exhibitions, Events, Admission, Correction, More, None, Finished\\
A: \ I was wondering what the hours are.\\
Q: \ We are open every day from 9:00 a.m. to 7:00 p.m.\\
KW: 9am, 10am, 11am, 12pm, 1pm, 2pm, Correction, More, None, \underline{Finished}\\
A: \ Thank you, good bye.\\
----------\\
\\
(EVAL) Make a reservation for 6 people at the pizzeria for Thursday at 7pm under the name\\
'Oliver'.\\
Q: \ Pizzeria romano how can i help you\\
KW: Delivery, Pickup, Menu, Hours, \underline{Reservations}, Specials, Correction, More, None, Finished\\
A: \ I would like to make reservations.\\
Q: \ Okay what day is the reservation for\\
KW: Monday, Tuesday, Wednesday, \underline{Thursday}, Friday, Saturday, Correction, More, None, Finished\\
A: \ Thursday at 7:00.\\
Q: \ Okay and how many people\\
KW: 1, 2, 3, 4, 5, \underline{6}, Correction, More, None, Finished\\
A: \ There will be 6 people.\\
Q: \ Okay so thursday at 7:00 for 6 people is that correct\\
KW: \underline{Yes}, No, Thursday, 7:00, Six, Correct, Correction, More, None, Finished\\
A: \ Yes.\\
Q: \ And what's your name\\
KW: Anna, Mayer, Anna Mayer, David Mayer, Laura, \underline{Oliver}, Correction, More, None, Finished\\
A: \ This is Oliver.\\
Q: \ All right oliver see you then\\
KW: Bye, Sure, OK, Alright, Goodbye, Yes, Correction, More, None, \underline{Finished}\\
A: \ Thank you, good bye.\\
----------\\
\\
(EVAL) Call your fitness studio and schedule a session with your trainer Sophia for Friday\\
morning under the name David Mayer.\\
KW: Join, Membership, Schedule, Classes, Instructors, Hours, Correction, \underline{More}, None, Finished\\
KW: Pricing, Locations, \underline{Trainers}, Services, Equipment, Programs, Correction, Previous, None,\\
Finished\\
A: \ I want to know more about the trainers.\\
Q: \ Okay which trainer in particular are you looking for\\
KW: Anna, Mayer, Anna Mayer, David Mayer, Laura, Oliver, Correction, \underline{More}, None, Finished\\
KW: Peter, \underline{Sophia}, Tim, Marianne, Maria, Felix, Correction, Previous, None, Finished\\
A: \ Sophia is the one I'm interested in.\\
Q: \ Would you like to make an appointment with her\\
KW: \underline{Yes}, No, Maybe, Soon, Later, Now, Correction, More, None, Finished\\
A: \ Yes, please.\\
Q: \ And which day works best for you\\
KW: Monday, Tuesday, Wednesday, Thursday, \underline{Friday}, Saturday, Correction, More, None, Finished\\
A: \ Friday.\\
Q: \ Okay do you prefer morning or afternoon\\
KW: \underline{Morning}, Afternoon, Early, Late, Dawn, Dusk, Correction, More, None, Finished\\
A: \ Morning.\\
Q: \ It's 10:00 a.m. okay\\
KW: \underline{Yes}, No, 10:00, 10am, 10 a, 10am ok, Correction, More, None, Finished\\
A: \ Yes.\\
Q: \ All right and what's your name\\
KW: Anna, Mayer, Anna Mayer, \underline{David Mayer}, Laura, Oliver, Correction, More, None, Finished\\
A: \ David Mayer.\\
Q: \ I've noted down your appointment\\
KW: Yes, No, Right, Correct, Accurate, Wrong, Correction, More, None, \underline{Finished}\\
A: \ Thank you, good bye.\\
----------\\
\\
(EVAL) Make a doctor's appointment for next Thursday morning under the name 'Anna Mayer'.\\
Q: \ Practice doctor garcia how can i help you\\
KW: \underline{Appointment}, Prescription, Test, Referral, Treatment, Diagnosis, Correction, More, None,\\
Finished\\
A: \ I need to make an appointment.\\
Q: \ Okay what's your name\\
KW: Anna, Mayer, \underline{Anna Mayer}, David Mayer, Laura, Oliver, Correction, More, None, Finished\\
A: \ My name is Anna Mayer.\\
Q: \ And which day would work best for you\\
KW: Monday, Tuesday, Wednesday, \underline{Thursday}, Friday, Saturday, Correction, More, None, Finished\\
A: \ Thursday would be fine.\\
Q: \ Okay and it's 9:00 a.m. okay\\
KW: \underline{Yes}, No, Correct, Incorrect, 9:00, 09:00, Correction, More, None, Finished\\
A: \ Yes.\\
Q: \ Okay perfect see you then\\
KW: Bye, See you, See ya, Later, Take care, Ciao, Correction, More, None, \underline{Finished}\\
A: \ Thank you, good bye.\\
----------\\
\\
(EVAL) Make a reservation at the cafe for 5 people next Tuesday at 5pm under the name Laura.\\
Q: \ Cafe marie how can i help you\\
KW: Order, Menu, Takeout, Delivery, Payment, \underline{Reservations}, Correction, More, None, Finished\\
A: \ I'd like to make a reservation.\\
Q: \ What day would you like to make the reservation for\\
KW: Monday, \underline{Tuesday}, Wednesday, Thursday, Friday, Saturday, Correction, More, None, Finished\\
A: \ Tuesday, please.\\
Q: \ And at what time are you coming\\
KW: 6am, 7am, 8am, 9am, 10am, 11am, Correction, \underline{More}, None, Finished\\
KW: 12pm, 1pm, 2pm, 3pm, 4pm, \underline{5pm}, Correction, Previous, None, Finished\\
A: \ 5:00 PM.\\
Q: \ Okay and how many people\\
KW: 1, 2, 3, 4, \underline{5}, 6, Correction, More, None, Finished\\
A: \ 5, please.\\
Q: \ What's your name\\
KW: Anna, Mayer, Anna Mayer, David Mayer, \underline{Laura}, Oliver, Correction, More, None, Finished\\
A: \ This is Laura.\\
Q: \ All right laura see you\\
KW: Bye, Goodbye, See you, See ya, Have a good day, Take care, Correction, More, None, \underline{Finished}\\
A: \ Thank you, good bye.\\
----------\\
\\
(EVAL) Order a taxi to 28 Bumblebee Lane for Monday at 5pm for 5 people.\\
Q: \ Taxi service jones how can i help you\\
KW: Book, Cancel, Inquiry, \underline{Reservation}, Schedule, Status, Correction, More, None, Finished\\
A: \ I'd like to make a reservation.\\
Q: \ Okay for today\\
KW: Yes, \underline{No}, Sure, Absolutely, Of course, Certainly, Correction, More, None, Finished\\
A: \ No, I'd like to make a reservation for next week.\\
Q: \ Okay which day\\
KW: \underline{Monday}, Tuesday, Wednesday, Thursday, Friday, Saturday, Correction, More, None, Finished\\
A: \ Monday for the week.\\
Q: \ Okay and what's your address\\
KW: 15 Flowerstreet, 7 Southroad, 104 Mainstreet, \underline{28 Bumblebee Lane}, 56 Park Avenue,\\
7 St. Michael Road, Correction, More, None, Finished\\
A: \ 28 Bumblebee Lane.\\
Q: \ And at what time would you like to be picked up\\
KW: 8:00am, 8:30am, 9:00am, 9:30am, 10:00am, 10:30am, Correction, \underline{More}, None, Finished\\
KW: 11:00am, 11:30am, 12:00pm, 12:30pm, 1:00pm, 1:30pm, Correction, Previous, \underline{None}, Finished\\
A: \ I am sorry, I cannot answer this question right now.\\
Q: \ I need to know the time at at when you would like to be picked up\\
KW: 8am, 9am, 10am, 11am, 12pm, 1pm, Correction, \underline{More}, None, Finished\\
KW: 2pm, 3pm, 4pm, \underline{5pm}, 6pm, 7pm, Correction, Previous, None, Finished\\
A: \ Around 5pm.\\
Q: \ Okay so monday at 5:00 p.m.\\
KW: Monday, Tuesday, Wednesday, Thursday, Friday, Saturday, Correction, \underline{More}, None, Finished\\
KW: Sunday, 5:00, 5:30, 6:00, 6:30, 7:00, Correction, \underline{Previous}, None, Finished\\
KW: \underline{Monday}, Tuesday, Wednesday, Thursday, Friday, Saturday, Correction, More, None, Finished\\
A: \ Okay, Monday at 5pm.\\
Q: \ And do you require an extra big taxi or is a regular car okay\\
KW: Big, Regular, \underline{Extra Big}, Small, Yes, No, Correction, More, None, Finished\\
A: \ I need an extra big taxi.\\
Q: \ All right i've got your reservation confirmed thanks\\
KW: Great, Perfect, Fantastic, Super, Wonderful, Excellent, Correction, More, None, \underline{Finished}\\
A: \ Thank you, good bye.\\
----------\\
\\
(EVAL) Make an appointment at the hair salon to have your hair colored on Monday afternoon under the name Laura.\\
Q: \ Leslie's hair salon how can i help you\\
KW: Directions, Hours, \underline{Appointment}, Services, Prices, Reviews, Correction, More, None, Finished\\
A: \ I have an appointment.\\
Q: \ Okay what's your name\\
KW: Anna, Mayer, Anna Mayer, David Mayer, \underline{Laura}, Oliver, Correction, More, None, Finished\\
A: \ Laura.\\
Q: \ I don't have an appointment for laura would you like to make one\\
KW: \underline{Yes}, No, Maybe, Later, Now, Tomorrow, Correction, More, None, Finished\\
A: \ Yes.\\
Q: \ For which day\\
KW: \underline{Monday}, Tuesday, Wednesday, Thursday, Friday, Saturday, Correction, More, None, Finished\\
A: \ Monday.\\
Q: \ Okay we could do 10:00 a.m. or 3:00 p.m. would one of those work for you\\
KW: 10am, \underline{3pm}, Yes, No, Maybe, Alright, Correction, More, None, Finished\\
A: \ I'll take the 3 p.m.\\
Q: \ Okay is the appointment for cutting coloring or styling\\
KW: Cutting, \underline{Coloring}, Styling, Both, Neither, All, Correction, More, None, Finished\\
A: \ Coloring.\\
Q: \ All right laura see you then\\
KW: Bye, See you, Bye bye, Take care, Good bye, See ya, Correction, More, None, \underline{Finished}\\
A: \ Thank you, good bye.\\
}

\subsubsection*{Subject 5}
\texttt{\frenchspacing
\noindent(TRAIN) Training scenario. Please follow the experimenter's instructions.\\
Q: \ Hello how are you doing today\\
KW: \underline{Good}, Great, Fantastic, Poorly, Terrible, OK, Correction, More, None, Finished\\
A: \ I'm doing good.\\
Q: \ What is your favorite food\\
KW: Pizza, Pasta, Burger, Steak, Salad, Sushi, Correction, \underline{More}, None, Finished\\
KW: Tacos, Soup, Sandwiches, \underline{Curry}, Rice, Noodles, Correction, Previous, None, Finished\\
A: \ Curry.\\
Q: \ Now please choose correction\\
KW: Right, Correct, Accurate, Exact, Perfect, Incorrect, \underline{Correction}, More, None, Finished\\
A: \ I am sorry, I misspoke earlier.\\
Q: \ Oh okay what is your favorite food\\
KW: \underline{Pizza}, Tacos, Sushi, Pasta, Burgers, Steak, Correction, More, None, Finished\\
A: \ Pizza is my favorite food.\\
Q: \ Please select none now\\
KW: \underline{None}, Zero, Nonexistent, Not applicable, Not selected, Not chosen, Correction, More, None,\\
Finished\\
A: \ None.\\
Q: \ Please select none on the right side\\
KW: None, Zero, Nada, Not any, None at all, Not one, Correction, More, \underline{None}, Finished\\
A: \ I am sorry, I cannot answer this question right now.\\
Q: \ Please enter scenario now by selecting finished on the right side\\
KW: Yes, No, Finished, Not yet, Almost, Ready, Correction, More, None, \underline{Finished}\\
A: \ Thank you, good bye.\\
----------\\
\\
(UITEST) Ask what the opening hours of the museum are.\\
Q: \ Museum of modern art how can i help you\\
KW: Visit, \underline{Hours}, Directions, Exhibits, Admission, Gift shop, Correction, More, None, Finished\\
A: \ What are the museum's hours?\\
Q: \ From 9:00 a.m. to 7:00 p.m.\\
KW: 9am, 10am, 11am, 12pm, 1pm, 2pm, Correction, More, None, \underline{Finished}\\
A: \ Thank you, good bye.\\
----------\\
\\
(EVAL) Make a reservation for 2 people at the pizzeria for Saturday at 1pm under the name\\
'Mayer'.\\
Q: \ Pizzeria romano how can i help you\\
KW: Order, Delivery, Pickup, Menu, Hours, Location, Correction, More, None, \underline{Finished}\\
A: \ Thank you, good bye.\\
----------\\
\\
(EVAL) Call your fitness studio and schedule a session with your trainer Felix for Monday\\
afternoon under the name David Mayer.\\
Q: \ Fitness studio get fit how can i help you\\
KW: Join, Price, Classes, Schedule, \underline{Trainers}, Location, Correction, More, None, Finished\\
A: \ We need trainers to help us get the word out.\\
Q: \ Okay are you a member\\
KW: Yes, \underline{No}, Maybe, Possibly, Certainly, Absolutely, Correction, More, None, Finished\\
A: \ No, I'm not.\\
Q: \ Would you like to sign up\\
KW: Yes, No, Absolutely, Definitely, Certainly, Maybe, \underline{Correction}, More, None, Finished\\
A: \ I am sorry, I misspoke earlier.\\
Q: \ Okay so how can i help you we are a fitness studio get fit\\
KW: \underline{Lose weight}, Gain muscle, Improve endurance, Increase flexibility, Tone body, Get stronger,\\
Correction, More, None, Finished\\
A: \ I'm looking to lose weight.\\
Q: \ Okay what do you like to schedule a private session\\
KW: Today, Tomorrow, \underline{This week}, Next week, Next month, Saturday, Correction, More, None,\\
Finished\\
A: \ This week.\\
Q: \ Okay sure which day\\
KW: Sunday, \underline{Monday}, Tuesday, Wednesday, Thursday, Friday, Correction, More, None, Finished\\
A: \ Any Monday will work for me.\\
Q: \ Okay do you already have a trainer\\
KW: \underline{Yes}, No, Maybe, Not yet, Not sure, Definitely, Correction, More, None, Finished\\
A: \ Yes, I do.\\
Q: \ All right what's your trainer's name\\
KW: Anna, Mayer, Anna Mayer, David Mayer, \underline{Laura}, Oliver, Correction, More, None, Finished\\
A: \ Laura.\\
Q: \ Okay on monday she is free at 10:00 a.m. or 3:00 p.m. would one of those work\\
KW: 10am, \underline{3pm}, Yes, No, Maybe, Possibly, Correction, More, None, Finished\\
A: \ Okay, the 3:00 p.m. one would be great.\\
Q: \ Okay perfect what's your name\\
KW: Anna, \underline{Mayer}, Anna Mayer, David Mayer, Laura, Oliver, Correction, More, None, Finished\\
A: \ Mayer.\\
Q: \ I need your full name please\\
KW: Anna, Mayer, Anna Mayer, \underline{David Mayer}, Laura, Oliver, Correction, More, None, Finished\\
A: \ David Mayer.\\
Q: \ Okay david see you then\\
KW: Bye, See you, Later, Goodbye, \underline{Hasta luego}, Adiós, Correction, More, None, Finished\\
A: \ Hasta luego.\\
Q: \ Bye-bye\\
KW: Bye, See ya, Farewell, Adios, Ciao, Hasta la vista, \underline{Correction}, More, None, Finished\\
A: \ I am sorry, I misspoke earlier.\\
Q: \ Okay what is your name\\
KW: Anna, Mayer, Anna Mayer, \underline{David Mayer}, Laura, Oliver, Correction, More, None, Finished\\
A: \ David Mayer.\\
Q: \ Okay and your trainer's name\\
KW: Anna, Mayer, Anna Mayer, \underline{David Mayer}, Laura, Oliver, Correction, More, None, Finished\\
A: \ Laura.\\
Q: \ And is monday at 3:00 p.m. working for you\\
KW: Yes, No, Monday, Tuesday, Wednesday, \underline{Thursday}, Correction, More, None, Finished\\
A: \ Thursday would be great.\\
Q: \ Okay so thursday 3:00 p.m.\\
KW: Thursday, \underline{3pm}, 3:00, 3:00pm, 15:00, 15:00pm, Correction, More, None, Finished\\
A: \ Okay, 3:00 p.m. is perfect.\\
Q: \ And it's thursday also perfect\\
KW: Yes, No, Right, Wrong, Absolutely, Certainly, Correction, More, \underline{None}, Finished\\
A: \ I am sorry, I cannot answer this question right now.\\
Q: \ Which day would you prefer to come\\
KW: \underline{Monday}, Tuesday, Wednesday, Thursday, Friday, Saturday, Correction, More, None, Finished\\
A: \ Monday would be great.\\
Q: \ Okay so monday at 3:00 with your trainer laura is that correct\\
KW: Yes, \underline{No}, Monday, 3:00, Laura, Correct, Correction, More, None, Finished\\
A: \ No, with my trainer Laura.\\
Q: \ I'm sorry i don't know how to help you\\
KW: OK, That's alright, It's ok, No worries, No problem, Don't worry, Correction, \underline{More}, None,\\
Finished\\
KW: I understand, That's fine, I appreciate it, No need, Not necessary, Not a problem,\\
Correction, Previous, \underline{None}, Finished\\
A: \ I am sorry, I cannot answer this question right now.\\
Q: \ Would you like to go ahead with an appointment on monday at 3:00 with your trainer laura\\
KW: Yes, \underline{No}, Monday, 3:00, Laura, Sure, Correction, More, None, Finished\\
A: \ No, with my trainer Laura.\\
Q: \ Okay\\
KW: Yes, \underline{No}, Sure, Alright, Okay, Yup, Correction, More, None, Finished\\
A: \ No, I would like to go ahead with an appointment on thursday at 3:00 p.m.\\
Q: \ Okay sounds good\\
KW: Yes, No, Alright, Sure, Good, Fine, Correction, \underline{More}, None, Finished\\
KW: Great, Sounds Good, Not Bad, Agreed, Okay, Not Sure, Correction, \underline{Previous}, None, Finished\\
KW: Yes, No, Alright, Sure, \underline{Good}, Fine, Correction, More, None, Finished\\
A: \ Okay, thank you.\\
Q: \ Okay goodbye\\
KW: \underline{Bye}, Bye-bye, Farewell, See ya, Adios, Have a good day, Correction, More, None, Finished\\
A: \ Bye.\\
Q: \ Bye\\
KW: Goodbye, Bye, \underline{See ya}, Ciao, Adios, Peace, Correction, More, None, Finished\\
A: \ See ya.\\
Q: \ Okay bye\\
KW: \underline{Bye}, Goodbye, See you, See ya, Later, Adios, Correction, More, None, Finished\\
A: \ Bye.\\
----------\\
\\
(EVAL) Make a doctor's appointment for next Tuesday morning under the name 'Anna Mayer'.\\
Q: \ Practice doctor garcia how can i help you\\
KW: \underline{Ask}, Explain, Request, Inquire, Suggest, Clarify, Correction, More, None, Finished\\
A: \ I have a question.\\
Q: \ Okay would you like to ask the doctor in person\\
KW: Yes, No, Maybe, Ask, Request, \underline{Speak}, Correction, More, None, Finished\\
A: \ Can I just speak to him?\\
Q: \ You will need to make an appointment\\
KW: Call, \underline{Schedule}, Arrange, Book, Make, Set, Correction, More, None, Finished\\
A: \ How do I schedule one?\\
Q: \ You can schedule one with me would you like to do that\\
KW: \underline{Yes}, No, Maybe, Certainly, Certainly not, Absolutely, Correction, More, None, Finished\\
A: \ Yes, I would.\\
Q: \ Okay which day would work best for you\\
KW: Monday, \underline{Tuesday}, Wednesday, Thursday, Friday, Saturday, Correction, More, None, Finished\\
A: \ Can I come in on Tuesday?\\
Q: \ Yes tuesday works how about 10:00 a.m.\\
KW: Yes, No, OK, Sure, \underline{Agree}, Disagree, Correction, More, None, Finished\\
A: \ That sounds like a good time.\\
Q: \ Okay what's your name\\
KW: Anna, Mayer, \underline{Anna Mayer}, David Mayer, Laura, Oliver, Correction, More, None, Finished\\
A: \ My name is Anna Mayer.\\
Q: \ Okay perfect see you then\\
KW: Bye, See ya, See you, Goodbye, Later, Adios, Correction, More, None, \underline{Finished}\\
A: \ Thank you, good bye.\\
----------\\
\\
(EVAL) Make a reservation at the cafe for 4 people next Wednesday at 4pm under the name Tim.\\
Q: \ Cafe marie how can i help you\\
KW: Order, Menu, Delivery, Takeout, \underline{Reservations}, Pay, Correction, More, None, Finished\\
A: \ I would like to make a reservation.\\
Q: \ What time would you like to make a reservation for\\
KW: 5pm, 6pm, 7pm, 8pm, 9pm, 10pm, Correction, \underline{More}, None, Finished\\
KW: 11pm, 12am, 1am, 2am, 3am, 4am, Correction, \underline{Previous}, None, Finished\\
KW: 5pm, 6pm, 7pm, 8pm, 9pm, 10pm, Correction, More, \underline{None}, Finished\\
A: \ I am sorry, I cannot answer this question right now.\\
Q: \ Which day would you like to come by\\
KW: Monday, Tuesday, \underline{Wednesday}, Thursday, Friday, Saturday, Correction, More, None, Finished\\
A: \ I would like to come on Wednesday.\\
Q: \ Okay and what time\\
KW: 12PM, 2PM, 4PM, 6PM, 8PM, 10PM, Correction, \underline{More}, None, Finished\\
KW: \underline{12AM}, 2AM, 4AM, 6AM, 8AM, 10AM, Correction, Previous, None, Finished\\
A: \ 12:00 AM.\\
Q: \ Okay and how many people\\
KW: 1, 2, 3, 4, 5, \underline{6}, Correction, More, None, Finished\\
A: \ Six people.\\
Q: \ Okay and what's your name\\
KW: Anna, \underline{Mayer}, Anna Mayer, David Mayer, Laura, Oliver, Correction, More, None, Finished\\
A: \ Mayer.\\
Q: \ Okay so a table for 6:00 on wednesday at\\
KW: 6PM, 6:00, Wednesday, 6PM Wednesday, 6:00 Wednesday, 6PM Wed, \underline{Correction}, More, None,\\
Finished\\
A: \ I am sorry, I misspoke earlier.\\
Q: \ Okay which which time did you say you want to come\\
KW: 8am, 9am, 10am, 11am, 12pm, 1pm, Correction, \underline{More}, None, Finished\\
KW: 2pm, 3pm, \underline{4pm}, 5pm, 6pm, 7pm, Correction, Previous, None, Finished\\
A: \ 4:00 PM.\\
Q: \ All right and how many people\\
KW: 1, 2, 3, \underline{4}, 5, 6, Correction, More, None, Finished\\
A: \ 4 people.\\
Q: \ Okay and your name please\\
KW: Anna, Mayer, Anna Mayer, David Mayer, Laura, Oliver, Correction, \underline{More}, None, Finished\\
KW: Peter, Sophia, \underline{Tim}, Marianne, Maria, Felix, Correction, Previous, None, Finished\\
A: \ My name is Tim.\\
Q: \ Okay see you on wednesday tim\\
KW: Wednesday, Thursday, Friday, Saturday, Sunday, Monday, Correction, More, None, \underline{Finished}\\
A: \ Thank you, good bye.\\
----------\\
\\
(EVAL) Order a taxi to 15 Flowerstreet for Saturday at 9am for 2 people.\\
Q: \ Taxi service jones how can i help you\\
KW: Book, \underline{Schedule}, Reservation, Cancel, Price, Location, Correction, More, None, Finished\\
A: \ I need a taxi to pick me up at 3 pm.\\
Q: \ Okay on which day\\
KW: Monday, Tuesday, Wednesday, Thursday, Friday, \underline{Saturday}, Correction, More, None, Finished\\
A: \ Saturday.\\
Q: \ Okay do you require an extra big taxi or is a regular car okay\\
KW: Big, \underline{Regular}, Extra Big, Small, Yes, No, Correction, More, None, Finished\\
A: \ Regular car is okay.\\
Q: \ All right and what's your address\\
KW: 15 Flowerstreet, 7 Southroad, 104 Mainstreet, \underline{28 Bumblebee Lane}, 56 Park Avenue,\\
7 St. Michael Road, Correction, More, None, Finished\\
A: \ 28 Bumblebee Lane.\\
Q: \ Okay so saturday at 3:00 is that correct\\
KW: Yes, No, Saturday, 3:00, Correct, Incorrect, \underline{Correction}, More, None, Finished\\
A: \ I am sorry, I misspoke earlier.\\
Q: \ Okay what time would you like to be picked up\\
KW: 8am, 8:30am, \underline{9am}, 9:30am, 10am, 10:30am, Correction, More, None, Finished\\
A: \ I would like to be picked up at 9 am.\\
Q: \ Okay so saturday at 9:00\\
KW: Saturday, 9:00, \underline{Yes}, No, Sure, Absolutely, Correction, More, None, Finished\\
A: \ Yes.\\
Q: \ Okay perfect see you then\\
KW: See you, Bye, Got it, Sure, Yup, Bye bye, \underline{Correction}, More, None, Finished\\
A: \ I am sorry, I misspoke earlier.\\
Q: \ Would you like to be picked up at a different address\\
KW: \underline{Yes}, No, Maybe, Possibly, Certainly, Absolutely, Correction, More, None, Finished\\
A: \ Yes, it's 28 Bumblebee Lane.\\
Q: \ Yes i have that or what or is it a different address if so what is\\
KW: \underline{Yes}, No, Different, Same, Address, Street, Correction, More, None, Finished\\
A: \ Yes, it's 28 Bumblebee Lane.\\
Q: \ What is the address that you would like to be picked up at\\
KW: \underline{15 Flowerstreet}, 7 Southroad, 104 Mainstreet, 28 Bumblebee Lane, 56 Park Avenue,\\
7 St. Michael Road, Correction, More, None, Finished\\
A: \ It's 15 Flowerstreet.\\
Q: \ Okay 15 flower street bend see you on saturday\\
KW: Yes, No, OK, 15, Flower, Street, Correction, More, None, \underline{Finished}\\
A: \ Thank you, good bye.\\
----------\\
\\
(EVAL) Make an appointment at the hair salon to have your hair styled on Friday afternoon under\\
the name Laura.\\
Q: \ Leslie's hair salon how can i help you\\
KW: \underline{Book}, Schedule, Prices, Directions, Hours, Services, Correction, More, None, Finished\\
A: \ I would like to book an appointment.\\
Q: \ Okay what day would you like to come by\\
KW: Monday, Tuesday, Wednesday, Thursday, \underline{Friday}, Saturday, Correction, More, None, Finished\\
A: \ Can I come in on Friday?\\
Q: \ Yes of course would 9:00 a.m. work or maybe 4:00 p.m.\\
KW: 9am, \underline{4pm}, Yes, No, Maybe, Certainly, Correction, More, None, Finished\\
A: \ That's perfect. Thank you\\
Q: \ Which one 9:00 a.m. or 4:00 p.m.\\
KW: 9:00 a, 9:00, 9, 4:00 p, \underline{4:00}, 4, Correction, More, None, Finished\\
A: \ 4:00 p.m.\\
Q: \ Okay great is the appointment for cutting coloring or styling\\
KW: Cutting, Coloring, \underline{Styling}, Both, Neither, All, Correction, More, None, Finished\\
A: \ Styling.\\
Q: \ All right and what's your name\\
KW: Anna, Mayer, Anna Mayer, \underline{David Mayer}, Laura, Oliver, Correction, More, None, Finished\\
A: \ My name is David Mayer.\\
Q: \ Okay david see you then\\
KW: Yes, No, See you, Bye, Okay, Sure, \underline{Correction}, More, None, Finished\\
A: \ I am sorry, I misspoke earlier.\\
Q: \ Is your name not david\\
KW: Anna, Mayer, Anna Mayer, David Mayer, \underline{Laura}, Oliver, Correction, More, None, Finished\\
A: \ No, it's Laura.\\
Q: \ Okay laura see you then\\
KW: Yes, No, Bye, See you, Later, Cya, Correction, More, None, \underline{Finished}\\
A: \ Thank you, good bye.
}

\end{document}